\documentclass[desactivate]{aa}  

\usepackage{graphicx}
\usepackage{txfonts}
\usepackage[]{hyperref}
\usepackage{xcolor}
\usepackage{soul}
\usepackage{booktabs}
\usepackage{tabularx}
\usepackage{breqn}

\newcommand{\Rer}{\ensuremath{R_{\oplus}}}
\newcommand{\Mer}{\ensuremath{M_{\oplus}}}
\newcommand{\Msun}{\ensuremath{M_{\odot}}}
\newcommand{\Rpl}{\ensuremath{R_{\rm pl}}}
\newcommand{\Mpl}{\ensuremath{M_{\rm pl}}}
\newcommand{\Rroc}{\ensuremath{R_{\rm roche}}}
\newcommand{\Teq}{\ensuremath{T_{\rm eq}}}
\newcommand{\Feuv}{\ensuremath{F_{\rm EUV}}}

\newcommand{\Reff}{\ensuremath{R_{\rm eff}}}

\newcommand{\dyncm}{\ensuremath{\rm dyn\,cm^{-2}}}

\begin{document} 

   %\title{Employing Neural Networks to Improve the Quality of Grid-based  Atmospheric Mass Loss Predictions}
   %\title{Employing Neural Networks to Extract Grid-based Exoplanet Atmospheric Mass Loss Predictions}
   \title{Escape of Water- and Metal-enriched Atmospheres from compact Hot mini-Neptunes with CHAIN}
   \subtitle{}
   \titlerunning{Water- and Metal-enriched Atmospheres with CHAIN}
   \author{
   Daria Kubyshkina\inst{1,2}\thanks{JAE and CP-G have contributed equally to this study.} $^{\href{https://orcid.org/0000-0001-9137-9818}{\includegraphics[scale=0.5]{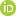}}}$,
   Jo Ann Egger\inst{3,1}\thanks{ESA research fellow} $^{\href{https://orcid.org/0000-0002-8275-1371}{\includegraphics[scale=0.5]{orcid.jpg}}}$,
   Caroline Piaulet-Ghorayeb\inst{4}\thanks{E. Margaret Burbidge Postdoctoral Fellow}$^{\href{https://orcid.org/0000-0003-4426-9530}{\includegraphics[scale=0.5]{orcid.jpg}}}$   
   }
  
   \authorrunning{Kubyshkina et al.}
   
   \institute{
    Space Research and Planetary Sciences, Physics Institute, University of Bern, Gesellschaftsstrasse 6, 3012 Bern, Switzerland \and
    Space Research Institute, Austrian Academy of Sciences, Schmiedlstrasse 6, A-8042 Graz, Austria \and
    European Space Agency (ESA), European Space Research and Technology Centre (ESTEC), Keplerlaan 1, 2201 AZ Noordwijk, The Netherlands \and
    Department of Astronomy \& Astrophysics, University of Chicago, 5640 South Ellis Avenue, Chicago, IL 60637, USA \\
    \email{daria.kubyshkina@unibe.ch}    
   }

   \date{Received XXX; accepted XXX}

    % \abstract{}{}{}{}{} 
    % 5 {} token are mandatory
 
   \abstract
  % leave it empty if necessary
   {Recent observations reveal that atmospheric compositions of close-in sub-Neptunes are diverse and can differ strongly from pure hydrogen-helium-dominated.}
   {We assess the possibility of modelling metal-rich and water-rich atmospheres with CHAIN model. We evaluate the major differences between the upper atmosphere photochemistry of such atmospheres compared to pure hydrogen-helium, and the impact on atmospheric mass loss rates. }
   {We employ CHAIN to model upper atmospheres of two warm and hot sub-Neptune-like planets which were suggested to host possibly water-rich or metal-rich atmospheres: GJ\,9827\,d and TOI-238\,b. For each planet, we consider a range of compositions spanning from H/He atmospheres with solar metal abundances to highly metal- and/or water-rich atmospheres and evaluate how our predictions change with increasing metal/water fractions.}
   {We find that for considered hot sub-Neptunes (1) an increase in water/metal abundance leads both to the increase in atmospheric cooling rates (due to the metal line cooling and molecular cooling processes) and the increase in heating (metal line heating in dense atmospheric layers and metals' photoionisation); (2) due to the increase in cooling and the mean particle weight of the atmosphere, the atmospheric mass loss drops significantly at high water/metal fractions (water mass fractions of 50\% or more or metal enrichment over 100 times solar), while at low enrichment levels mass loss rates are similar to those of H-He atmospheres with solar abundances or slightly higher; (3) for the same atmospheric mean particle weight, the escape from water-rich atmospheres is generally lower.}
   {In the context of atmospheric evolution, it implies that the atmospheres with initially high water fraction in the atmosphere are more stable. Furthermore, due to the preferential escape of hydrogen, the atmospheric metal/hydrogen ratio is expected to increase significantly with time, leading to the earlier cessation of the hydrodynamic escape.}

   
   \maketitle
%
%-
\section{Introduction}\label{sec:intro} 
Recent observations (in particular, the results of James Webb Space Telescope, JWST) indicate that the atmospheres of sub-Neptune-like planets are diverse in compositions and can differ strongly from nebula-like hydrogen-helium (H-He) compositions with stellar-like metallicities. 
For some planets, observational signatures suggest the presence in the atmospheres of water vapour in significant amounts \citep[e.g.,][]{Benneke2019NatAs...3..813B,Benneke2019ApJ...887L..14B,Brande2022AJ....164..197B,Fu2022ApJ...940L..35F,Piaulet2024ApJ...974L..10P,egger2024A&A...688A.223E,Hu2025arXiv250712622H,Murphy2025AJ....169..286M}, which is generally expected from the recent formation models for planets formed outside of the ice line \citep[e.g.][]{venturini2020,venturini2024A&A...686L...9V,Burn2024NatAs...8..463B,Burn2024arXiv241116879B}. Furthermore, some Neptune-like planets' atmospheres were found to have metallicities significantly higher than those of the host star
\citep[e.g.][]{Benneke2024arXiv240303325B,Ahrer2025ApJ...985L..10A}, which can be explained either by the specifics of their formation or the fractionation of the light and heavy elements during the evolution. 

This poses a challenge for modelling the evolution of planetary atmospheres using standard models designed for H/He-dominated atmospheres, as the escape rates depend on the atmosphere's composition \citep[e.g.][]{johnstone2015,GMunoz2024Icar..41516080G,Yoshida2022ApJ...934..137Y,Yoshida2024PEPS...11...59Y,Yoshida2025A&A...696L..13Y}. Furthermore, due to the fractionation, the atmospheric composition may not remain constant throughout the evolution \citep[e.g.][]{hunten1987,lammer2020,lammer2025NatAs...9.1022L,Malsky2020ApJ...896...48M,gu2023ApJ...953L..27G,Louca2023ApJ...956L..19L,Louca2025arXiv250506013L,Cherubim2024ApJ...967..139C}. 

On the other hand, it offers a possibility to explain poorly understood parts of mass-radius and radius-period distributions of low-mass exoplanets. In particular, this applies to {those} hot low-mass planets {residing near or in the hot Neptune desert or radius gap \citep[][]{fulton2017}, where observed radii are too large to be consistent with pure rocky composition} \citep[e.g.][referred to as E25 below]{Egger2025A&A...696A..28E}. For such {hot low-mass} planets, {structure modelling might suggest a presence of an extended atmosphere \citep[e.g.][]{Lopez_Fortney2014ApJ...792....1L,tang2025ApJ...989...28T,2025arXiv251201805P}, while} atmospheric evolution modelling repeatedly suggests that the pure H/He atmospheres are fully eroded by hydrodynamic escape within the first $\sim1-100$\,Myr \citep[e.g.][E25]{kubyshkina2019a,bonfanti2021evol,egger2024A&A...688A.223E,Leonardi2025arXiv250914156L}, through {blow-off/}boil-off {\citep[catastrophic escape powered through the bolometric heating of planetary atmosphere by the young star][]{Erkaev2015,owen_wu2016ApJ...817..107O,Ginzburg2016ApJ...825...29G,Tang2024ApJ...976..221T}}. This makes it hard to explain the presence of extended atmospheres at Gyr-old planets. On the other hand, high molecular weight atmospheres (1) are expected to be more stable against atmospheric escape {due to the reduction of the relevant atmospheric scale heights \citep[e.g.][]{tang2025ApJ...989...28T} and enhanced radiative cooling \citep[e.g.][]{WangDai2019ApJ...873L...1W,GMunoz2024Icar..41516080G,Yoshida2022ApJ...934..137Y,Yoshida2024PEPS...11...59Y}} and (2) correspond to larger volatile budget for a given set of planetary parameters \citep[i.e., for a typical sub-Neptune like planet, its mass and radius can be reproduced with internal structure modelling assuming the atmospheric mass fraction of either a few percent of planetary mass for H/He atmospheres, or a few tens percent for high molecular weight atmospheres, see e.g.][]{zeng2021gap_waterEOS,egger2024A&A...688A.223E,aguichine2025ApJ...988..186A}. Therefore, they can possibly survive on Gyr timescales even {at low mass planets} under extreme irradiation conditions.

{Distinguishing between different atmospheric compositions} can be complicated when using internal structure models alone (even employing proper equations of state and precise mass and radius measurements), as their solutions are often degenerate and the same planet parameters can be explained by a wide range of compositions \citep[e.g.][{E25}]{Valencia2007ApJ...665.1413V,rogers_seager2010,Kite2021ApJ...909L..22K,Piaulet2023NatAs...7..206P,egger2024A&A...688A.223E,Leonardi2025arXiv250914156L}. The atmosphere observations that could solve this degeneracy \citep[e.g.][]{Benneke2012ApJ...753..100B}, in turn, are not always feasible (e.g. too weak signal to noise ratio or high stellar variability, or limited observational time on suitable instruments). In such cases, analysing the atmosphere's stability employing a theoretical model can be a useful tool to disentangle the degeneracy \citep[e.g.][]{Piaulet2023NatAs...7..206P}. Such an approach is most effective for hot, close-in planets, where atmospheric escape is expected to be strongest.

To perform an analysis of the atmospheric stability in the evolution context, one needs to start from understanding the upper atmosphere's dynamics for various atmospheric compositions, planetary parameters and high-energy irradiation backgrounds. In this study, we assess the effect of the enrichment of H-He atmospheres in water and heavy elements for two highly-irradiated sub-Neptune-like planets using Cloudy e Hydro Ancora INsieme (CHAIN) model, employing a highly detailed (photo)chemical reactions network.
We describe our modelling approach and the target planets in Sec.\,\ref{sec:model}. In Sec.\,\ref{sec:results}, we describe the outputs of our simulations and discuss the contributions of specific processes (\ref{sec:results_heating_and_cooling}--\ref{sec:results_low_irradiation}) and their impact of the bulk atmospheric parameters (\ref{sec:results_mass_loss_etc}). We discuss the possible influence of the assumptions of our models and the implications of our results in a context of planetary evolution in Sec.\,\ref{sec:discuss} and summarise our findings in Sec.\,\ref{sec:conclusions}.
%
%%%%%%%%%%%%%%%%%%%%%%%%%%%%%%%%%%%%%%%%%%%%%%%%%%%%%%%%%%%
%\section{Composition of hot sub-Neptune-like planets}\label{sec:SN_composition}%
%
%\subsection{Insights from observations}\label{sec:SN_composition_observations} %
%
%\subsection{Insights from internal structure modelling}\label{sec:SN_composition_internal}%
%

\section{Modelling approach}\label{sec:model}
\subsection{Upper Atmosphere Model}\label{sec:model_chain}
%-
The Cloudy e Hydro Ancora INsieme (CHAIN) code combines the one-dimensional (1D) hydrodynamic model of the escaping planetary upper atmosphere \citep[][]{kubyshkina2018grid} with the 1D hydrostatic non-local thermodynamic equilibrium (NLTE) photochemical and radiative transfer solver Cloudy \citep[versions C17, C23, see ][respectively]{ferland2017,Chatzikos2023RMxAA..59..327C}, following the approach first introduced in \citet{salz2015}. The detailed description of the model along with the assessment of the model's performance across a wide parameter range can be found in \citet{kubyshkina2024cloudy} (further referred to as K24). In brief, the hydrodynamic code solves the fluid dynamic equations, while calculations of the heating and cooling processes included in the energy conservation equation, along with the chemical state of the atmospheric elements (atomic, molecular, ion, and excited states), are delegated to the Cloudy code. The Cloudy code accounts for elements {with atomic numbers between 1 and 30 (hydrogen to zinc)} with adjustable relative abundances, their ions and a wide range of excited states, and 122 molecules including water\footnote{{C17 employs the data from https://home.strw.leidenuniv.nl/~moldata/ \citep[][]{Schoeier2005A&A...432..369S}}}. The hydrodynamic code considers the atmosphere as one-fluid, with the mean particle weight set by the considered atmospheric composition. The stellar input is provided in the form of stellar spectra, where we scale separately the X-ray (wavelength $\lambda<10$\,nm), extreme ultraviolet (EUV, 10-91.2\,nm) and ultraviolet (UV) to infrared (IR) parts of the spectra to fit the desired values, which can be adopted from observations or stellar models.
\subsection{Application for metal- and water-rich atmospheres}
\label{sec:model_enriched}
By default, the abundances of different elements in Cloudy are set at solar composition. Each element's abundance can be adjusted separately by introducing an enrichment (depletion) factor relative to solar abundances, which changes the element content uniformly throughout the considered gas medium, or prescribing the altitude-dependent element fractions relative to hydrogen. As the hydrodynamic code used in CHAIN considers a one-fluid atmosphere, in the present study we use the first option.

For metal-enriched atmospheres, we scale the elemental abundances relative to hydrogen uniformly for all elements between lithium and zinc and consider enrichment between $\times10-500$ of solar values. For water-rich atmospheres, the approach is less straightforward. Per se, Cloudy code does not allow to fix the molecular abundances at the lower boundary of the simulation domain (or anywhere in the atmosphere); one can only adjust the elemental abundances, while their state is decided by the photochemistry framework. Therefore, we assume that {the atmospheric mixture consists of H/He gas with solar metallicity, and the added water vapour is assumed to be well mixed within that H/He background, which is }
%} and (2) {the water molecules are fully photodissociated in most of the simulation domain}. Both assumptions are 
expected to be reasonable for the upper atmospheres (pressures {$\lesssim1000$\,\dyncm}) of the considered planets, where temperatures are generally {close to 1000\,K or higher} \citep[][]{Haldemann2020A&A...643A.105H}. We then translate the water mass fraction ($f^{\rm m}_{\rm H_2O}$) to the number ratio of water molecules to the hydrogen atoms in the H-He part of the atmosphere, and adjust the elemental abundances of oxygen to hydrogen ($Z_{\rm O}$) and of hydrogen to all other elements in the gas accordingly. {We note, that here and further in the paper we use Z and $Z_{\rm O}$ as metals and oxygen number ratio to hydrogen rather than general metallicity.} The resulting relations between water mass fraction, $Z_{\rm O}$, and the mean particle weight of the atmosphere ($\mu$) are summarised in Tab.\,\ref{tab:fH2O-ZO-mu}. {This directly sets the composition in the uppermost atmospheric regions, where most of the molecules are photodissociated. In lower layers (pressures above $\sim$1-10\,\dyncm), the} resulting abundances of different molecules %(mostly stable at pressures above 1-10\,\dyncm) 
are then set by the (photo)chemical framework of Cloudy {(see details in Sec.\ref{sec:results_h-he}--\ref{sec:results_metals})}. {Under conditions considered in this study, the most abundant molecules are H$_2$, H$_2$O, O$_2$, CO, SiO, N$_2$, and NO.} {Both for the metal-rich and water-rich atmospheres, He fraction relative to H remains at the solar value default for Cloudy.} 
\begin{table}[]
    \centering
    \caption{Relation between the water mass fraction, oxygen enrichment {(number ratio to hydrogen)}, and the mean molecular weight of the atmosphere.}
    \begin{tabular}{c|c|c}
    \toprule
      $f^{\mathrm m}_{\mathrm H_2O}$   &  $Z_{\mathrm O}/Z^{\odot}_{\mathrm O}$ &  $\mu\, [m_{\mathrm H}]$\\
      \midrule
       0.1  & 13.67 & 1.40 \\
       0.3  & 50.77 & 1.76 \\
       0.5  & 111.1 & 2.38 \\
       0.7  & 226.4 & 3.64 \\
       0.8  & 335.0 & 4.96 \\
       0.9  & 534.5 & 7.78 \\
%       0.95  & 713.4 & 10.1 \\
       \bottomrule
    \end{tabular}    
    \label{tab:fH2O-ZO-mu}
\end{table}

We note that the Cloudy framework, as well as our hydrodynamic code, are built for the hydrogen-dominated atmospheres, i.e. a gas where hydrogen flow defines the atmospheric dynamics. With the gradual increase of the relative abundance of heavy elements to hydrogen, at some point the latter condition is not satisfied anymore and the applicability of our model comes in question. Therefore, we limit the discussion to the atmospheres with the relative heavy elements abundances up to $Z = 500\,Z_{\odot}$ and water mass fraction up to 90\%, where hydrogen makes up at least $\sim50\%$ of the atmospheric atoms. %We also focus on planets and conditions where hydrogen flow can still drag heavier elements along, at least in the lower parts of the upper atmosphere \citep[which we verify a posteriori using the approach described in][]{Catling2017aeil.book.....C}. 
%We note, that we also performed a few simulations for higher heavy element abundances and did not observe any qualitative changes in the outputs; however, we refrain from discussing these results here and only include them as supplementary information in Tables\,\ref{tab:gj9827d_simlist0}-\ref{tab:toi238b_simlist1}.
%
Finally, for reference, we simulate the atmospheres of {GJ\,9827\,d and TOI-238\,b} for H-He composition with solar heavy element abundance and, for GJ\,9827\,d, for pure H-He and pure H atmospheres.

\subsection{Planetary parameters and model settings}\label{sec:model_planets}
We consider two planets for which we previously made probe runs with our model: a warm sub-Neptune with confirmed water vapour in the atmosphere GJ\,9827\,d \citep[][]{Piaulet2024ApJ...974L..10P} and potential hot water-world TOI-238\,b (i.e. the planet hosting an atmosphere in the region where H-He atmospheres with {solar}-like metallicities are not stable; see {E25}). Both planets are compact (radii \Rpl$\leq2$\,\Rer) and have masses ($M_{\rm pl}$) about $\sim3.5$\,\Mer\ and orbit moderately active K-dwarf stars, but GJ\,9827\,d has a radius about 25\% larger than TOI-238\,b, while the latter orbits closer to its star and has an equilibrium temperature (\Teq) about twice higher than GJ\,9827\,d. We summarise planetary parameters in Table\,\ref{tab:planets}.
\begin{table}[]
    \centering
    \caption{Planetary parameters adopted for GJ\,9827\,d and TOI-238\,b.}
    \begin{tabular}{c|c|c|c|c|c|c}
    \toprule
        planet &  \Teq & \Rpl & \Mpl & $M_*$ & \Feuv & a \\
            &   [K] & [\Rer] & [\Mer] & [\Msun] & [$\frac{erg}{s\,cm^2}$] & [AU] \\
        \midrule
        GJ\,9827\,d & 610 & 2.0 & 3.5 & 0.606 & 2923 & 0.055 \\ 
        TOI-238\,b & 1270 & 1.6 & 3.4 & 0.79  & 13836 & 0.0256 \\
        \bottomrule        
    \end{tabular}   
    \footnotesize{Parameters for GJ\,9827\,d and TOI-238\,b were adopted from \citet{Bonomo2023A&A...677A..33B} and \citet{Mistry2024PASA...41...30M,SuarezMascareno2024A&A...685A..56S}, respectively. Temperatures \Teq were set as described in K24.}
    \label{tab:planets}
\end{table}

{We are most interested in early-time evolution of the atmospheres, as most consequential for atmospheric evolution. Therefore, to maximize the irradiation and atmospheric mass loss rates,} we adopt the X-ray and EUV irradiation levels typical for young ($\sim500$\,Myr) stars according to the predictions of the Mors stellar evolution code \citep[][]{johnstone2021mors}, assuming that the star evolved as fast rotator. \footnote{{This assumption can be valid, even though stars appear quite moderate among other K-dwarfs \citep[][]{Carleo2021AJ....161..136C,SuarezMascareno2024A&A...685A..56S}, due to their potentially old (though poorly constrained) ages, because the evolution of different stellar rotators cannot be distinguished for stars older than 1-2\,Gyr. However, even if this assumption is not true, the planets would remain highly irradiated due to their extreme proximity to the host stars. We investigate a few cases with reduced irradiation in Sec.\,\ref{sec:results_low_irradiation}.}}
%. At this time, the stars remain in the saturation regime; therefore, the effect of different stellar rotation rates is irrelevant. 
For both stars we employ the spectrum (SED) of GJ\,436 \citep[from Measurements of the Ultraviolet Spectral Characteristics of Low-mass Exoplanetary Systems, MUSCLES, survey][]{musclesI,musclesII,musclesIII,musclesIV,musclesV}.%, which is different from the approach we previously used in \citet{Piaulet2024ApJ...974L..10P} and \citet{Egger2025A&A...696A..28E}.

As the Cloudy framework is limited to the total densities below $10^{15}$\,cm$^{-3}$ (and for denser atmospheric parts we extend Cloudy predictions at constant values), we set our lower boundary close to this limit. As our default lower boundary pressure, we adopt {500-1000\,\dyncm}, and for a few cases, we additionally consider {pressures of 10000 and 100\,\dyncm} to test the effect of this condition on our results. {We note, that assigning the observed radius to the arbitrarily set, non-photospheric, pressure, might not be strictly accurate. However, an inaccuracy introduced by this assignment is expected to be minor as the difference in radial distances corresponding to the atmospheric pressure range {1000-100,000 \dyncm} is typically small \citep[e.g.][]{kubyshkina2024handbook}}. The temperature at the lower boundary is set to the black-body equilibrium temperature of the planet. By default, we set the upper boundary at 1.5 {times} the Roche radius of the planet (\Rroc) for GJ\,9827\,d and at 1.8\Rroc\ for TOI-238\,b and apply the continuous outflow conditions there. 

For GJ\,9827\,d, we further consider a few cases with the planet on more distant orbits/receiving less X-ray and/or EUV radiation (for the H-He atmosphere with solar metal abundance, referred to as H-He atmosphere in what follows for brevity, and for the water-rich atmosphere with $f^{\rm m}_{\rm H_2O} = 0.5$), as detailed in Sec.\,\ref{sec:results}. More details on the input parameters and the list of simulations for GJ\,9827\,d and TOI-238\,b can be found in Tab.\,\ref{tab:gj9827d_simlist0} and Tab.\,\ref{tab:toi238b_simlist0}, respectively.

\begin{figure}
    \centering
    \includegraphics[width=0.7\linewidth]{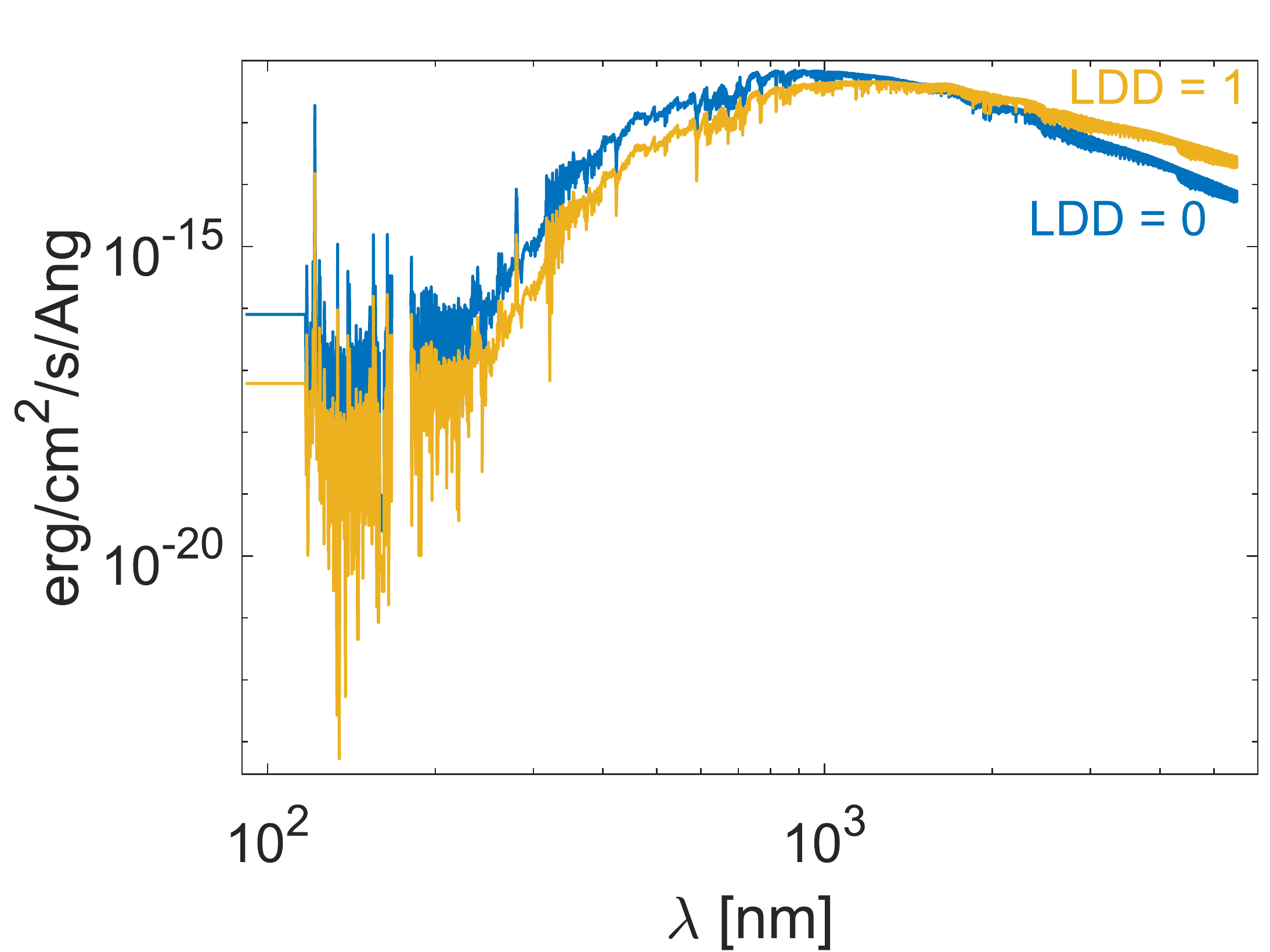}
    \caption{The shape (unscaled) of the default (blue) and ``IR-enriched'' spectra used in this study.}
    \label{fig:ldd0_to_ldd1}
\end{figure}

To better understand the features found in our simulations ({see} Sec.\,\ref{sec:results_heating_and_cooling}) {and test their robustness against the model assumptions}, we considered two additional model configurations along with the default settings described above and in K24. First, {observations suggest that the shape of the spectra is not universal among stars of the same type, not only in terms of the local (line) features, but also in terms of the ratio of IR to visible and UV radiation, even if the total and XUV radiation is similar (see e.g. the references to the MUSCLES survey above). Therefore,} along with the default spectrum of GJ\,436, we run simulations with the spectral energy at wavelengths $\lambda>91.2$\,nm redistributed towards longer wavelengths, as illustrated in Fig.\,\ref{fig:ldd0_to_ldd1}. To achieve this, we multiply the spectrum by the linear function ($\sim\lambda$) and re-scale the spectrum to keep the total luminosity constant ($\simeq L_{\rm bol}$ of the given star); X-ray and EUV luminosities are kept constant. We refer to this type of spectrum as ``IR-enhanced'' throughout the paper, but it can also be considered as UV-visible light depleted. 
%In model names in Tables\,\ref{tab:gj9827d_simlist0}-\ref{tab:toi238b_simlist1} the usage of this spectrum is noted as ``LDD\,=\,1'', while the default spectrum corresponds to ``LDD\,=\,0''.
%

Furthermore, we performed runs disabling some parts of Cloudy's photochemical framework. Namely, we switched off the Ly$\alpha$ pumping (fluorescent excitation) of the H-like iso-sequence (marked as ``NOLyaP'' in Tables\,\ref{tab:gj9827d_simlist0}-\ref{tab:toi238b_simlist1}, only for H-He atmospheres) or all of the induced processes (marked as ``NOIND'' in Tables\,\ref{tab:gj9827d_simlist0}-\ref{tab:toi238b_simlist1}, for all atmospheres except pure H/H-He without heavier elements). We explain these adjustments in more detail in Sec.\,\ref{sec:results_heating_and_cooling}. In total, we performed 63 simulation for GJ\,9827\,d and 51 simulation for TOI-238\,b.
%Similar analysis was performed also for TOI-1685\,b, with mass and radius similar to those of TOI-238\,b and irradiation level in between of the two considered planets; however, as the results are close to those of TOI-238\,b, we do not include them here.

%
\section{Modelling results}\label{sec:results}
As described above, we performed simulations over a range of compositions with mean particle weight $\mu$ ranging between 1-7 m$_{\rm H}$. With increasing $\mu$, the density and pressure gradients in the atmospheres become steeper and most of the characteristic distances (such as photodissociation and ionisation fronts for different elements, sonic point and the exobase position) move closer to the planet. This leads to a strong drop in atmospheric densities at high altitudes (e.g., at \Rroc) {between the lowest and highest $\mu$ models}: up to about an order of magnitude for GJ\,9827\,d and up to 2 orders of magnitude for TOI-238\,b. Meanwhile, the bulk outflow velocity at the same distances increases by factors of 2 and 3, respectively, and remains close to constant with increasing $\mu$ for $\mu\gtrsim4$\,$m_{\rm H}$ for each atmospheric/model type. Consequently, the atmospheric mass loss rates ($\dot{M}$) drop, but at somewhat different rates for different atmospheric types and conditions. The changes in density gradients lead to the peak temperatures increasing with increasing $\mu$ (with a sort of saturation effect at high $\mu$), while the altitude of the thermosphere moves downwards (similarly to the changes accompanying the increase in planet's mass).

In the following, we describe in more detail the contribution of specific processes. We start with the overview of the heating and cooling processes %on the example of GJ\,9827\,d 
assuming a H-He atmosphere with (1) a solar metal abundance (further referred to as ``default'' composition), (2) a water-rich atmosphere with $f^{\rm m}_{\rm H_2O} = 0.7$, and (3) a heavy-element-enriched atmosphere (which we further refer to as ``metal-rich'') with $Z = 300\times Z_{\odot}$ and discuss the impact of various model parameters in Sec.\,\ref{sec:results_heating_and_cooling}. In Sec.\,\ref{sec:results_h-he}--\ref{sec:results_metals} we provide more details specific for different atmospheric compositions. Sec.\,\ref{sec:results_low_irradiation} reviews the results for the cases of low irradiation. We then discuss how some of the important atmospheric parameters change across all of the considered models in Sec.\,\ref{sec:results_mass_loss_etc}. {The underlying temperature and bulk velocity profiles for the models discussed below can be found in Appendix\,\ref{apx:summary_table}.}

\subsection{Heating and cooling of highly irradiated atmospheres}\label{sec:results_heating_and_cooling}
%
%\subsubsection{Radiative cooling}
%
\begin{figure}
    \centering
    \includegraphics[width=0.8\linewidth]{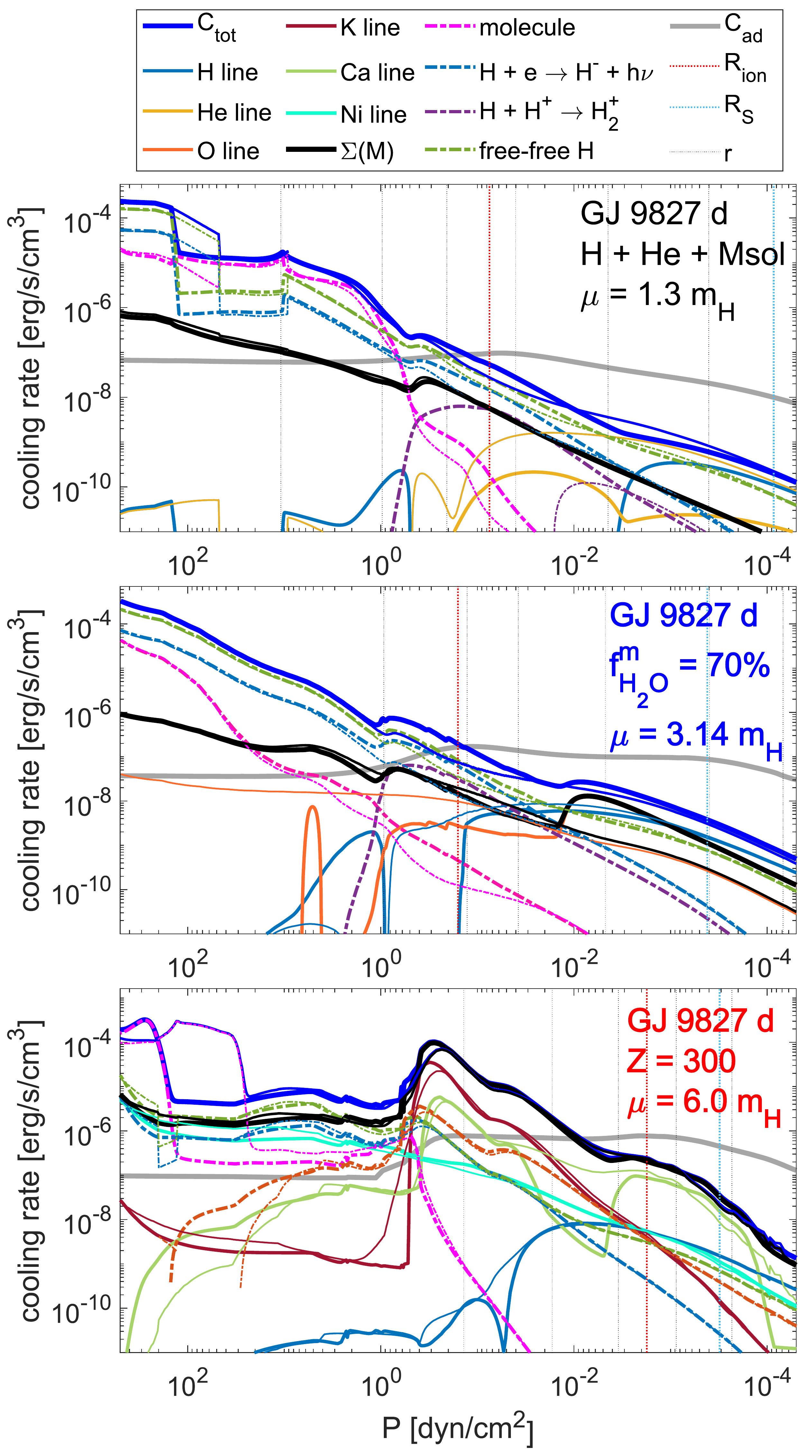}
    \caption{Radiative volume cooling rates for three different atmospheric types: H/He with solar metallicity (top), water-enriched with $f^{\rm m}_{\rm H_2O}\,=\,70\%$ (middle), and H/He with metallicity $\times$300 of the solar (bottom). Cooling processes included in the plots are shown in the legend; thicker and thinner lines of the same type show the models with and without induced processes, respectively. Molecular cooling does not include the effects from H$_2$. SED is default (LDD\,=\,0). The lines defined as r denote the radial distances of 1.1, 1.2, 1.3, 1.5, 2.0, and 3.0\,\Rpl.}
    \label{fig:cool_compn_noind}
\end{figure}
\begin{figure}
    \centering
    \includegraphics[width=0.7\linewidth]{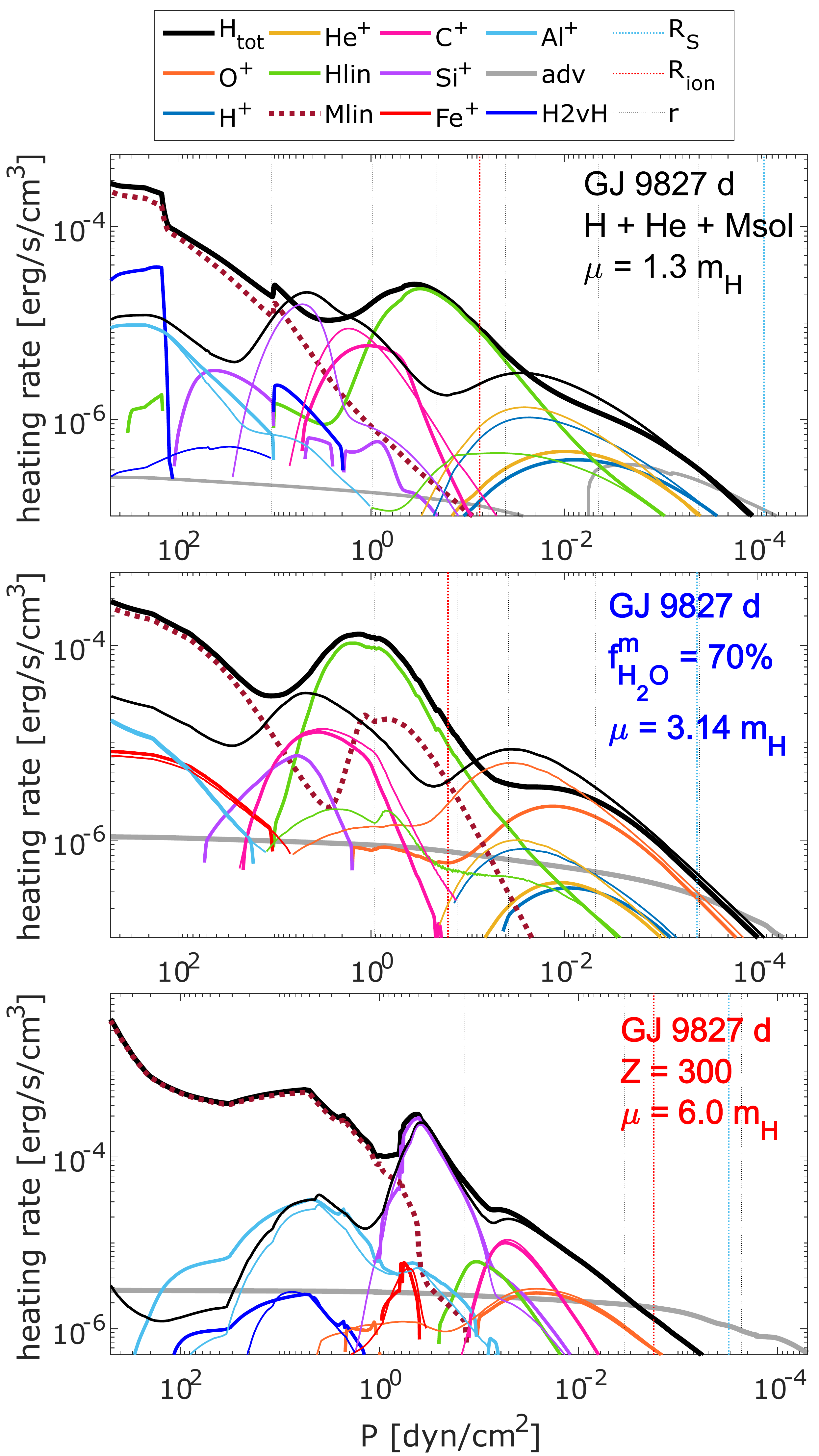}
    \caption{Volume heating rates for the same atmospheric types as in Fig.\,\ref{fig:cool_compn_noind}, as shown in the plots. Heating processes included in the plots are shown in the legend; thicker and thinner lines of the same type show the models with and without induced processes, respectively. H2vH term in the legend is the heating by collisional de-excitation of vibrationally excited H$_2$. SED is default (LDD\,=\,0). The lines defined as r denote the radial distances of 1.1, 1.2, 1.3, 1.5, 2.0, and 3.0\,\Rpl.}
    \label{fig:heat_compn_noind}
\end{figure}
%Volume heating rates and their most relevant contributors (>5% of Htot at each specific r) 
%The processes responsible for the heating and the corresponding line styles are given in the legend  and follow this notation: Htot – total volume heating rate from all sources; Hlin – iso-sequence line heating; Hi – heating due to photoionisation of atomic hydrogen; Hion – collisional ionisation heating (all hydrogen species); H- – H− -heating; Hn = 2 – photoionisation from all excited states of hydrogen species; H2ph – H2 photoionisation heating; H2vH – heating by collisional de-excitation of vibrationally excited H2; H2dH – heating by photodissociation of H2; Heii – photoionisation of Heii; Hei – photoionisation of Hei; ChaT – charge transfer heating; Caii – metal line heating dominated by absorption of the Caii line at 7291.47 Å; Oi – photoionisation of Oi; Fei – photoionisation of Fei.

In terms of the cooling processes, for all considered cases, radiative cooling ($C_{\rm rad}$) dominates at lower altitudes (pressures P above 0.1-10\,\dyncm), while higher altitudes are dominated by the cooling from adiabatic expansion. In Fig.\,\ref{fig:cool_compn_noind}, we show the volume cooling rates for most relevant RC processes (contributing $\gtrsim10\%$ to the total $C_{\rm rad}$ at some altitudes) for the default, water-rich, and metal-rich atmospheres.  {For the context, we also include the positions of the sonic points ($R_{\rm S}$) and ionisation maximum ($R_{\rm ion}$), as an equivalent of the effective radius of the photoionisation heating; we note, however, that, due to the complex photochemistry included in the models, the latter is not quite representative of the whole photoheating picture. }%, with the mean molecular weight increasing from the first to the last (top to bottom panels). 
At the lowermost altitudes ($P>1$\,\dyncm), most of the $C_{\rm rad}$ comes from molecular cooling (not accounting for H$_2${, which contributes, through the recombination of H$_2^+$, up to 9\% of $C_{\rm rad}$ near $\sim1$\,\dyncm}), free-free interaction of H (bremsstrahlung cooling), and production of H$^-$. The molecular cooling increases systematically for the IR-enhanced SED (up to the factor of $\sim3$). In case of the metal-rich atmosphere, significant contribution at these pressures comes from the line cooling associated with heavier elements, such as Mg, Fe, and Ni; for H-dominated and water-rich atmospheres, their contribution at these altitudes is an order of magnitude below that of molecular cooling. For H-dominated and water-rich atmospheres with lower $f^{\rm m}_{\rm H_2O}$, molecular cooling makes up to 100\% of total $C_{\rm rad}$ in this region (which is also about 5-10 times higher than in the case if we switch off the molecule formation in Cloudy's chemical framework). However, with an increasing $f^{\rm m}_{\rm H_2O}$, hence increasing $\mu$, the photodissociation front of H$_2$ and H$_2$O, which make up most of the molecules in the water-enriched atmospheres, move to lower altitudes, until it, essentially, reaches the lower boundary of our simulation domain at $f^{\rm m}_{\rm H_2O}\sim0.7-0.8$. This can have a twofold effect on our results (see Appendix\,\ref{apx:results_pressure}) and suggest that to fully assess the effect of molecular cooling on such atmospheres, we need to employ a model that can operate at higher densities.

At higher altitudes, the $C_{\rm rad}$ for H-He atmosphere is dominated by H-line and He-line cooling, which can be complemented by the various metal line cooling processes between 0.01-1\,\dyncm\,(in particular, C, Na, and K lines). For water-enriched atmospheres, the main contributions in this region come from H and O lines and recombination of H$^+_2$. In both cases, however, the $C_{\rm rad}$ at pressures below $\sim1$\,\dyncm\ is much lower than that of adiabatic expansion. For highly metal-enriched atmospheres, however, the total contribution from metal lines at 0.001-1\,\dyncm\ can reach values comparable to the molecule cooling near the lower boundary and make up to 100\% of both radiative and total cooling; furthermore, metal-line cooling remains a significant cooling constituent at $10^{-4}-10^{-3}$\,\dyncm\,(due to, e.g. Ca line).

%\subsubsection{Heating}
%
While the cooling processes are in line with what's expected based on the previous modelling performed in K24 and by other authors, we encounter a range of atypical processes when applying the standard configuration of Cloudy, which we did not see for more moderate planets (see Fig.\,\ref{fig:heat_compn_noind}). They are associated specifically with the induced processes. First, so-called hydrogen iso-sequence line heating (``Hlin'' in the legend), was briefly discussed in K24. It is associated with the heating due to the overabundance of the excited H in S2 and P2 states; specifically, through the following path: H(1S) + $h\nu(\gtrsim12{\,\rm eV}) \rightarrow$ H(2P) $\rightarrow$ H(2S) + $h\nu(\sim2{\,\rm eV})$ (fluorescent excitation). The radiative de-excitation in the last part of this chain contributes to cooling; further on, the resulting atom in the metastable 2S level can be photoionised, also by lower-energy photons, or lose its energy in the collisional de-excitation. For highly-irradiated atmospheres of planets considered here, Hlin becomes a main contributor at pressures between 0.01-0.1 and 1-10\,\dyncm in case of the H-He and water-rich atmospheres; for metal-rich atmospheres, its contribution remain 1-2 orders of magnitude below that of heavy ions. Testing confirms that the Hlin term is fully controlled by the stellar irradiation in VUV interval $\sim6.5-12.4$\,eV (in particular, Ly$\alpha$) and becomes irrelevant in its absence. The term ``Mlin'' (see the legend of Fig.\,\ref{fig:heat_compn_noind}) represents the sum of contributions from similar processes including heavier species; it is mostly controlled by lower-energy photons and dominates the heating at pressures above 1-10\,\dyncm for all compositions. In its absence, the heating in this region is about an order of magnitude lower for H-He and water-rich atmospheres and up to three orders of magnitude for the metal rich atmospheres. We also note, that despite the apparent prominence of the Mlin term, it is limited to the densest part of the upper atmospheres, which were shown to have a limited impact on the bulk outflow (see also Appendix\,\ref{apx:results_pressure}). The rest of the heating is comprised of photoionisation of hydrogen and helium (at pressures lower than $\sim$0.1\,\dyncm) and heavier species (at pressures above $\sim0.01$\,\dyncm). In Fig.\,\ref{fig:heat_compn_noind}, we omit the contributions from S which are similar to C, and from Mg, Ca, and Fe similar to Al to keep the plots readable. We note that in case of the IR-enhanced SED the heating from the photoionisation of heavy species (e.g., Si) at $\sim$0.1\,\dyncm\ is significantly reduced, while it remains similar at lower altitudes.

\begin{figure}
    \centering
    \includegraphics[width=0.6\linewidth]{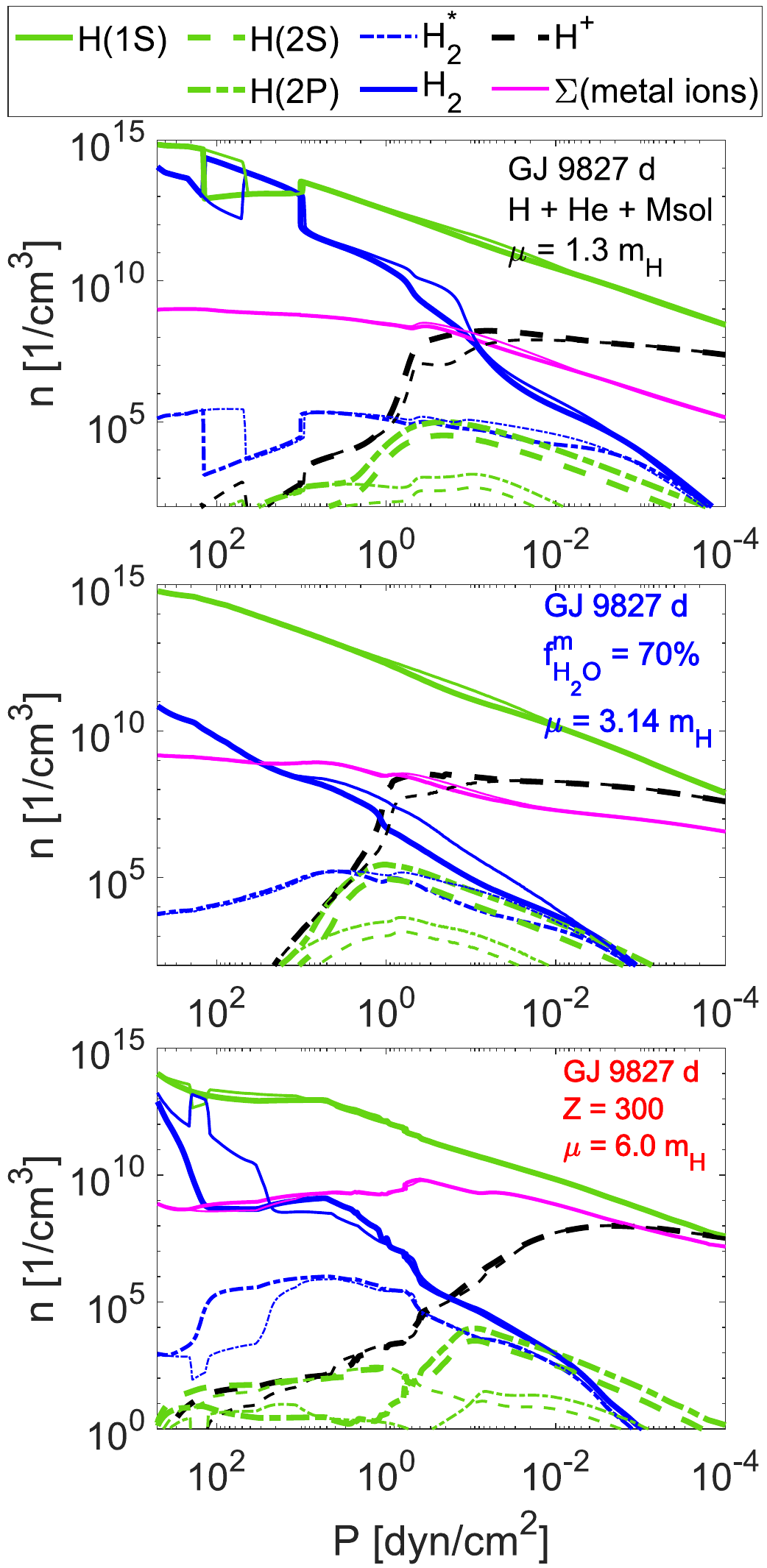}
    \caption{Populations of some hydrogen species and heavy ions for the same atmospheric types as in Fig.\,\ref{fig:cool_compn_noind}, as shown in the plots. Different line types are explained in the legend and thicker and thinner lines of the same type show the models with and without induced processes, respectively. SED is default (LDD\,=\,0).}
    \label{fig:hpops_compn_noind}
\end{figure}

%\subsubsection{Contributions from the excited species}
The prominence of Hlin and Mlin terms should be taken with caution for a range of reasons. First, particularly relevant for Mlin, pressures near the lower boundary of the simulation domain are near the applicability limit of Cloudy. The testing for a few chosen models did not revel qualitative changes in the model predictions with changes in the lower boundary position/pressure. We {first} tested pressures down to 100 times smaller than the default values in post-processing, by re-running Cloudy using the same final density-temperature profiles, {thus reassessing the heating and cooling processes; these remained at the same level for a given pressure. We then tested} pressures of 10 times smaller and larger than the default in the self-consistent simulations; {we found that increasing pressure leads to a little increase in the mass loss rates, density and temperature of the outflow and decreasing - vice versa (see details in Appendix\,\ref{apx:results_pressure}). However, as Cloudy model can not operate at high pressures, we note that } the lower boundary conditions {may} still have effects on our results {that can not be reproduced with the current framework}\footnote{The recent release of Cloudy \citep[][]{Gunasekera2025arXiv250801102G} includes a revision of molecular database in order to improve code's compatibility with the atmospheric studies and might provide more robust results. However, for the present study we use the Cloudy configurations which were previously tested within CHAIN.}. Furthermore, the opacity treatment in Cloudy might not account properly for the possible ion shielding (one can see in Fig,\,\ref{fig:heat_compn_noind} that photoionisation processes typically occur at higher altitudes, specifically for hydrogen), which can also lead to the overestimation of Hlin and Mlin \citep[e.g.][]{Kawamura2024ApJ...967...95K}. Finally, some species contributing to the Mlin term will likely not be present in realistic atmospheres due to the condensation at lower altitudes (we discuss this issue in more detail in Sec.\,\ref{sec:discuss_condensation}; {we note, that besides condensation effects, the wind advection is not treated fully self-consistently in the present model, whic can also be relevant, see Appendix\,\ref{apx:results_wind-adv}}). Therefore, along with the default configuration of the model, we decided to perform the same simulations but switching off the whole range of induced processes in Cloudy. The {result of such change for the fixed temperature and density is} shown by thinner lines of the same types in Figures\,\ref{fig:cool_compn_noind}-\ref{fig:hpops_compn_noind}. One can see that in terms of the cooling processes, little changes with switching off the induced processes, except that certain line cooling processes contribute at broader ranges of altitudes (specifically, H and O lines). For the heating processes, however, it leads to the Mlin term becoming neglectful and Hlin becoming a secondary contributor; instead, the contribution from photoionisation of H, He, and some heavier species (e.g., O, C) from the ground state, increases.

In Fig.\,\ref{fig:hpops_compn_noind}, we show the number density profiles of some hydrogen species and for the same cases as considered in Figures\,\ref{fig:cool_compn_noind} and \ref{fig:heat_compn_noind}. One can see, that in models without induced processes the density of hydrogen in 2S and 2P states is significantly lower than in models with induced processes at higher altitudes, while populations at lower altitudes are weakly affected. Similar effect can be seen for the excited molecular hydrogen near 100\,\dyncm, though less pronounced; this does not, however, has a clear effect on heating or cooling and the effect depends strongly on local conditions. For H-He atmosphere, the boundary is around 1\,\dyncm\ and it moves to $\sim10$\,\dyncm\ for the water-rich atmospheres; for metal-rich atmospheres, it remains at the similar level due to the extra heating from metals at low altitudes. The total ion density declines slowly with altitude ($\sim$50 times between 500 and $10^{-4}$\,\dyncm, in comparison with more than 6 orders of magnitude for H(1S) across the same pressures). Above $\sim10^{-3}-10$\,\dyncm\,(lowest value corresponds to the metal-enriched atmospheres and highest to the water-enriched), the ion population is dominated by H$^+$, while at lower pressures it is composed of various heavy ions. The transition between these two regimes roughly coincides with the peak in H(2s); we also see a decrease in H$^+$ densities with exclusion of the induced processes near this point.

For the IR-enhanced SED, the contribution of molecular cooling relative to other processes increases, leading to the generally higher $C_{\rm rad}$ at low altitudes. The total metal line cooling at higher altitudes increases for water-rich atmospheres but decreases at high metal-enrichment levels. The heating rates decline (specifically at higher altitudes) due to the irradiation being smaller in VUV/FUV wavelength range; this effect, however, becomes less significant for high enrichment levels.

The results for TOI-238\,b are qualitatively similar to the described above. One of the main differences is the somewhat weaker effect of induced processes for H-He and water-rich atmospheres: the enhancement in excited hydrogen states is below the factor of 50 and occurs at higher altitudes. For the high metal enrichment cases, however, it can be up to 3 orders of magnitude in certain regions. The abundance of ions and excited species for the hotter (and more EUV irradiated) planet is generally higher. The exclusion of the induced processes in this case leads not only to strong changes in heating but also to the significant decrease in the $C_{\rm rad}$ near the pressure levels of 10$^{-2}-1$\,\dyncm\ and a more pronounced increase in the heating from the photoionisation of heavier species (at similar altitudes as the decrease in cooling) for H-He and water-rich atmospheres than in the case of GJ\,9827\,d. For metal-rich atmospheres, exclusion of the induced processes leads to up to an order of magnitude decrease in cooling also at pressures above $\sim$1\,\dyncm and local increases in cooling at altitudes above $\sim$1\,\dyncm\ level; the total heating behaves similarly. We include the plots similar to those in Figures\,\ref{fig:cool_compn_noind}-\ref{fig:hpops_compn_noind} in the Appendix\,\ref{apx:extras} (Fig.\,\ref{fig:cool-heat-pop_toi238b}).

\subsection{H-He-dominated atmospheres}\label{sec:results_h-he}
In H-He atmospheres, most of the molecular budget is comprised by H$_2$ (which is the dominant hydrogen form at the lower boundary, but can also be present in significant amounts at high altitudes), with H$_2$O and CO being minor constituents (at low altitudes, $\sim10^{-3}$ compared to H$_2$ for GJ\,9827\,d and even lower for TOI-238\,b; number of CO molecules becomes about 1\% of that of H$_2$ near 1\,\dyncm). H$_2$O and CO have similar abundances at the lower boundary, but water is significantly less stable against photoevaporation. At high altitudes, the molecular budget is mostly comprised of light molecules; along with H$_2$ (including ionised and excited states), HeH$^+$ ions dominate at pressures lower than $10^{-4}-10^{-3}$\,\dyncm.
The cooling in H-He atmospheres is dominated by molecular cooling processes at low altitudes (notably, contribution from H$_2$O and CO is comparable to that of hydrogen despite their low abundances), and by hydrogen-related cooling processes at high altitudes (production of H$^-$, free-free interactions, recombination of H$^+_2$, and H-line cooling). Heating at high altitudes is also mostly controlled by hydrogen photochemistry; at low altitudes, we see a non-negligible contribution from metals for strongly irradiated atmospheres, which was, however, shown to have a limited effect on $\dot{M}$ due to the high atmospheric densities in this region. Results for the metal-free H-He atmospheres are briefed in Appendix\,\ref{apx:results_met-free}.

\subsection{Water-rich atmospheres}\label{sec:results_water}%
\begin{figure}
    \centering
    \includegraphics[width=0.6\linewidth]{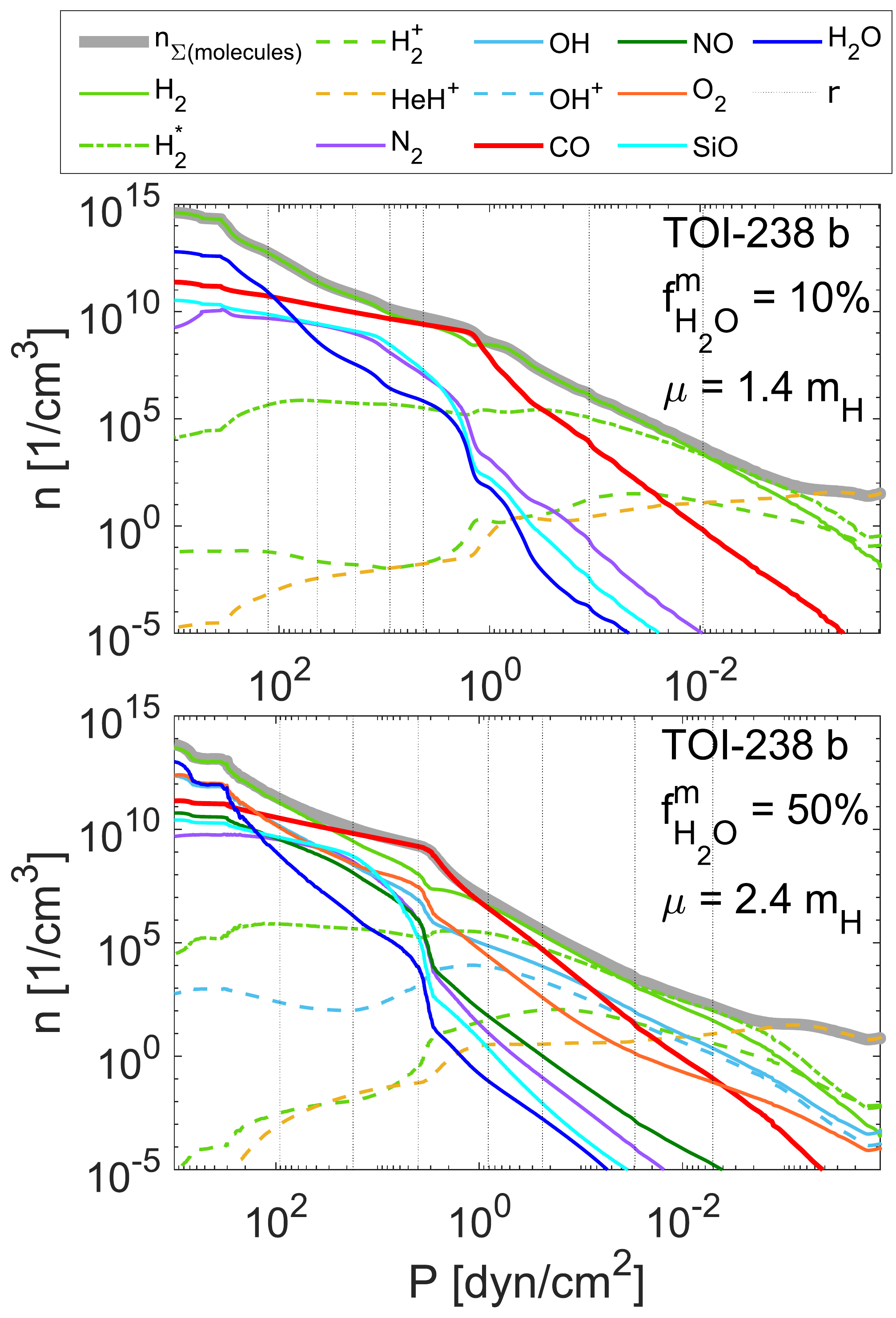}
    \caption{Number densities of molecule species in water-rich atmosphere models (default configuration) for TOI-238\,b. In top panel and bottom panels, $f^m_{H_2O} = 10\%$ and $f^m_{H_2O} = 50\%$, respectively. The lines are explained in the legend. The $n_{\Sigma(molecules)}$ term refers to the total molecular budget and the lines defined as r denote the radial distances of 1.1, 1.2, 1.3, 1.4, 1.5, 2.0, and 3.0\,\Rpl.}
    \label{fig:molecules_H2O}
\end{figure}
\begin{figure}
    \centering
    \includegraphics[width=0.61\linewidth]{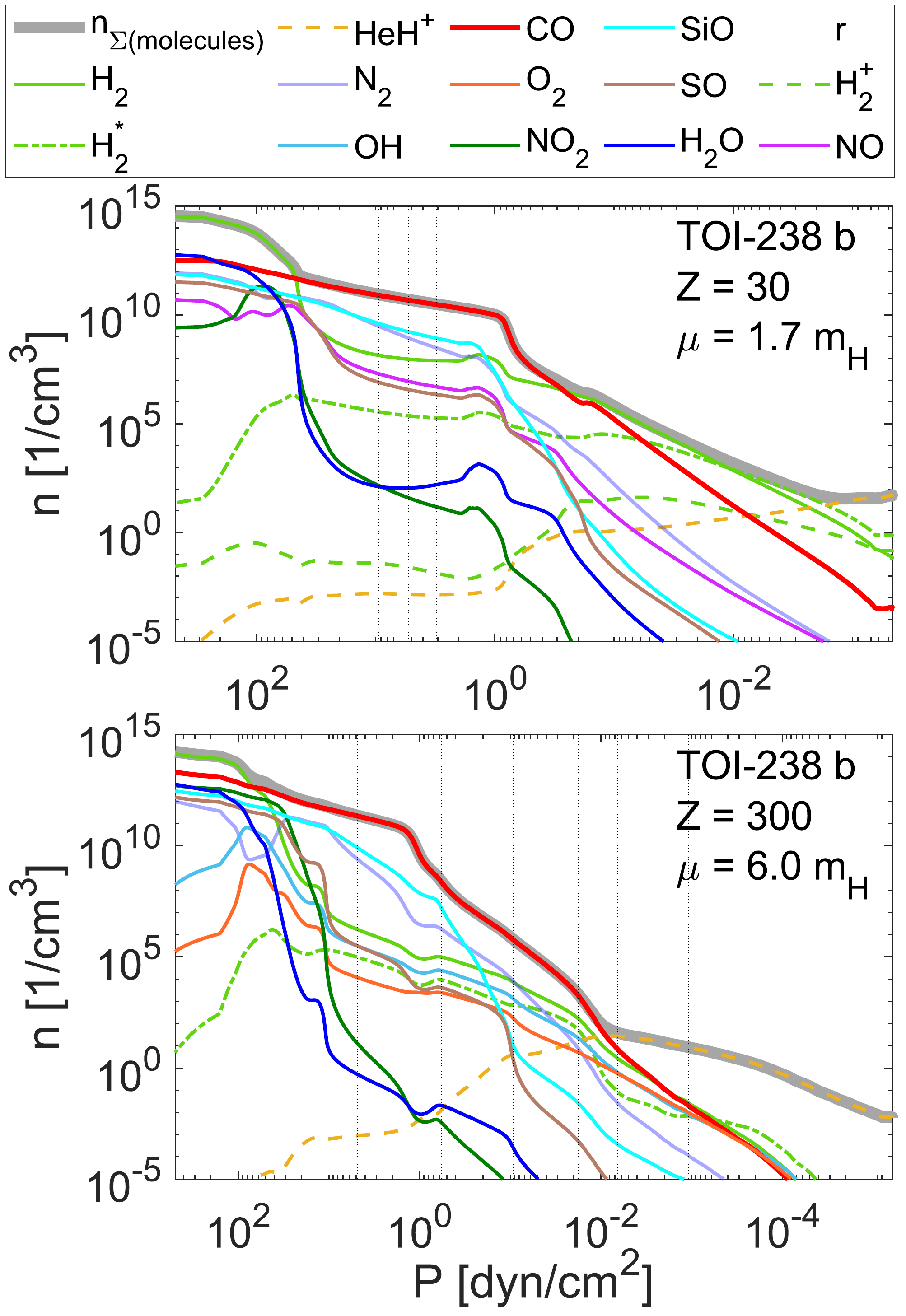}
    \caption{Same as in Figure\,\ref{fig:molecules_H2O}, but for metal enriched atmospheres. Top and bottom panels correspond to the metal enrichment 30 and 300 times of solar, respectively.}
    \label{fig:molecules_HM}
\end{figure}
In case of water-rich atmospheres, most radiative processes remain controlled by hydrogen, and qualitative picture is similar to the H-He atmospheres. At high altitudes ($P$ below $\sim$0.1-1\,\dyncm), oxygen line cooling becomes a significant contribution to the $C_{\rm rad}$; however, it occurs in the region where the total cooling is dominated by the adiabatic expansion. At high water mass fractions {(${f_{\rm H_2O}^{\rm m} > 0.5}$)}, photoionisation of oxygen (and secondary photoionisation in case of TOI-238\,b) contributes significantly to heating at high altitudes {(local contribution to the total heating rate reaches up to 10, 20, and 70\% for $f_{\rm H_2O}^{\rm m}$ of 0.5, 0.7, and 0.9, respectively; we note, however, that this contribution still does not exceed $\sim1.3$\% of the heating rate associated with hydrogen)}. 

The molecular content in the atmosphere becomes more diverse; in Figure\,\ref{fig:molecules_H2O}, we show two examples for a low ($f^m_{H_2O} = 10\%$) and a moderate ($f^m_{H_2O} = 50\%$) water mass fractions for the case of TOI-238\,b planet (for higher $f^m_{H_2O}$, the picture is similar; the picture is also qualitatively similar for GJ\,9827\,d). The ratio of $n_{\rm CO}/n_{\rm H_2}$ at the lower boundary is similar to the H-He case, while the water abundance relative to H$_2$ increases $\sim10-500$ times. However, due to the lower photodissociation potential, H$_2$O abundances drop below those of CO already around $\sim$100\,\dyncm, and CO remains the most abundant heavy molecule at high altitudes. Furthermore, the diversity of molecules with significant abundances relative to H$_2$ increases with the water mass fraction on account of oxygen-bearing species. At $f^m_{H_2O} = 10\%$, it includes SiO with the abundance similar to N$_2$; at $f^m_{H_2O} = 50\%$ and higher, it also includes NO (similarly abundant as SiO), O$_2$ (similarly abundant as H$_2$O at lower boundary but more stable against photodissociation), SO in the case of the cooler GJ\,9827\,d, and OH/ OH$^+$ at high altitudes. 

We note, that the assessment of the molecular abundances at the lower boundary of our simulation domain for water-rich atmospheres is complicated by the fact that with the increasing water enrichment (hence, $\mu$), and the consecutive increase in ion fraction at high altitudes, the photodissociation fronts for different molecules move {to higher pressures and lower altitudes}. As the position of our lower boundary is limited by the applicability limits of Cloudy, at some point the photodissociation fronts of H$_2$ and H$_2$O technically shift outside of the modelling domain for GJ\,9827\,d and water-rich atmospheres. The bulk atmospheric parameters proceed to change smoothly when it occurs (see Sec.\,\ref{sec:results_mass_loss_etc}) and shown no qualitative difference compared to the cases unaffected by this issue; therefore, we assume that our general results are weakly affected. We provide more detail in Appendix\,\ref{apx:results_pressure}. %However, some of the model predictions should be taken with caution, which we specify below.

\subsection{Metal-rich atmospheres}\label{sec:results_metals}
%-
At high metal abundances, the hydrogen-driven processes are no longer necessarily dominant in $C_{\rm rad}$ and heating of the highly-irradiated atmospheres. %We note, however, that the planets considered here present an extreme example and for cooler and less dense planets the effects considered here are much less pronounced (we performed some testing with CHAIN for TOI-469\,b, \citealt{egger2024A&A...688A.223E}, and GJ\,436\,b, e.g., \citealt{butler2004}) and should not be extrapolated on any generic planet. Furthermore, these results are not applicable to the actual secondary atmospheres, where hydrogen is not a dominant constituent of the atmosphere (even if they experience a hydrodynamic escape).
Differently to the water-rich atmospheres, where the metal (oxygen) line cooling is only relevant at relatively high altitudes, the total metal cooling in the metal-rich atmospheres is comparable or even higher than that of hydrogen-driven processes at low altitudes and dominates the cooling processes at pressures below $\sim0.1-1$\,\dyncm for metal abundances $Z\gtrsim50\times Z_{\odot}$ (for lower Z, it is smaller or comparable to H-line cooling). At the lowermost altitudes ($P$ above 10-100\,\dyncm), the cooling is dominated by molecular cooling processes (cooling rates about an order of magnitude higher than other processes), due to the increased molecule abundances. In terms of heating, metals contribute at low altitudes (through the Mlin term) in all cases, and at high altitudes (through the photoionisation of heavy species) at the metal enrichment of $\gtrsim50\times Z_{\odot}$ for GJ\,9827\,d and of $\gtrsim30\times Z_{\odot}$ for TOI-238\,b.

In Fig.\,\ref{fig:molecules_HM}, we show the molecule abundances for $Z = 30\times Z_{\odot}$ and $Z = 300\times Z_{\odot}$ for the case of TOI-238\,b. As in the case of the water-rich atmospheres, the results for GJ\,9827\,d are qualitatively similar but the molecule number densities are generally higher due to the lower atmospheric temperatures. Also in metal-rich atmospheres, H$_2$O and CO remain the most abundant molecules except H$_2$; their number ratio to H$_2$, however, increases by the factor of $\sim10$ for similar $\mu$ {relative to the water-rich case (the relative abundances are similar for both planets)}. Water and CO abundances at the lower boundary remain similar, as in the case of H-He atmospheres. The H$_2$O abundance, however, is smaller than in the water-rich atmospheres case for the same O/H enrichment; H$_2$O/CO ratio also slightly decreases with increasing $Z$. Both effects owe to the oxygen being relatively easily bound in other molecules when their counterparts are present in the atmosphere in sufficient amounts. The variety of molecules present in significant amounts is higher than for the water-rich atmospheres, even at the relatively low $Z$ values; along with the species discussed above, it also includes NO$_2$, but only in the case of TOI-238\,b. %We note, that the molecule abundances discussed above can be affected by the adiabatic wind advection (see Appendix\,\ref{apx:results_wind-adv}).

\subsection{The cases of low temperature and low high-energy irradiation}\label{sec:results_low_irradiation}
We further tested how the outputs of our models change with decreased irradiation for GJ\,9827\,d. Therefore, we ran the model for the reduced X-ray radiation (0.5, and 0.1 of the $F_{\rm X}$ in the default case), reduced XUV ($\sim0.3$ and $\sim0.2$, and $\sim0.15^*$ of the default case, referred to as ``XUV-5Gyr'', ``XUV-9Gyr'', and ``0.14AU'' in Tab.\,\ref{tab:gj9827d_simlist0}, respectively), and planet moved to 400$^*$\,K and 300$^*$\,K orbits (the respective changes in XUV are about 0.15$^*$ and 0.05$^*$ of the default). These tests we only applied for H-He atmospheres with solar metal abundances and water-rich atmospheres with $f^{\rm m}_{\rm H_2O} = 50\%$ where numbers are marked with ``$^*$''.

For low X-ray cases, we find no difference with the default case except {for the reduction in cooling caused by free-free H interactions and production of H$^-$ at pressures between 100-500\,dyn/cm$^2$. It occurs because of the decrease in the number of free electrons, presumably connected to the reduction in metal ions production in this pressure interval (most prominent for magnesium and iron). Meanwhile, metal line cooling in the same interval remains at the level similar to the default case.} It does not affect any bulk properties of the atmospheres significantly. For the cases where EUV is also reduced (reduced XUV), we find the decrease in mass loss rates roughly similar or less {than} the decrease in EUV flux for H-He atmospheres but about 6 times larger than the decrease in EUV for the water-rich case; decrease is also stronger for the IR-enhanced SED. The latter is due to the fact that the decrease in EUV only affect the H$^+$ heating and, to some extent, the Hlin term, while it affects less the contributions from heavier species. The Mlin term only decreases significantly for the water-rich case, which we associate with the increase of molecular abundances at the relevant altitudes. Finally, decrease in both \Teq\ and XUV leads to the significant drop in the escape rates (up to about two orders for H/He atmosphere and over a three orders of magnitude for the water-rich atmosphere); in this case, the Mlin term drops for both H-He and water-rich atmospheres.

We note that in some of the reduced XUV cases and in most of the cases with both \Teq\ and XUV reduced, the exobase position gets near or below the sonic point. This implies that the applicability conditions for our model are barely fulfilled or not fulfilled, respectively (this does not apply to the predictions of the photochemical framework of Cloudy, as they do not require the considered gas medium to be hydrodynamic). Therefore, to calculate the $\dot{M}$ (such cases are marked with asterisks in Table\,\ref{tab:gj9827d_simlist0}) we used the parameters predicted by our models below the exobase (where it is technically valid) and calculated the Jeans-like escape rates a posteriori. We note, however, that the obtained values are typically within a factor of 3 from the values predicted by the hydrodynamic model. Furthermore, for such atmospheres, the (atomic) hydrogen is expected to be the only escaping species, while for the majority of the simulations in this study, the outflow is strong enough to drag some heavier species along with hydrogen (see a more detailed discussion in Sec.\,\ref{sec:discuss}).
\begin{figure*}[ht]
    \centering
    \includegraphics[width=0.8\linewidth]{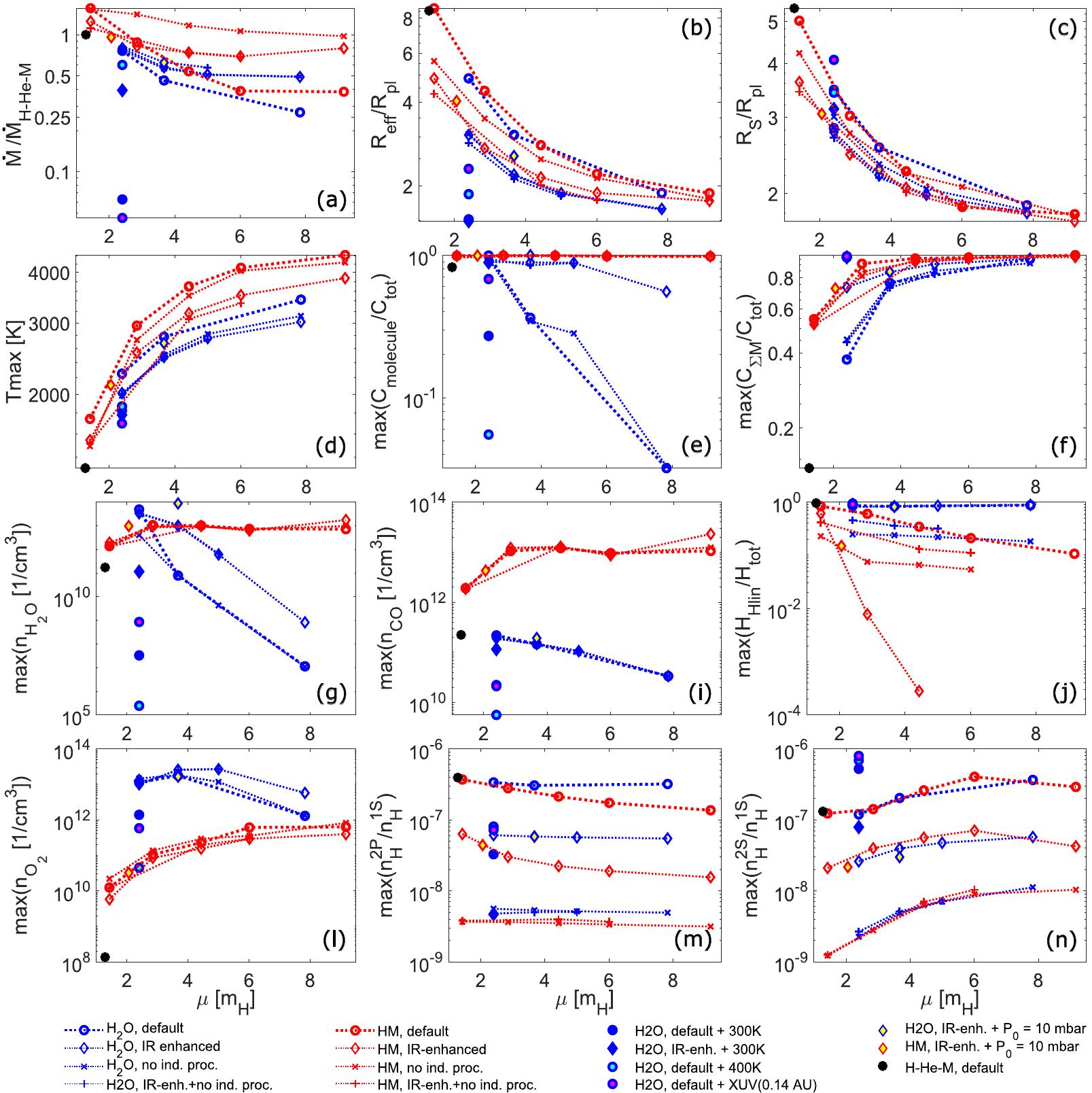}
    \caption{Summary of the simulation outputs for different model settings for GJ\,9827\,d: ratio of the mass loss rate to the default model (H + He + M_{\rm sol}, IR\,=\,0, ind\,=\,1) (a); indicative effective radius of the EUV absorption \Reff\,(b); sonic radius $R_{\rm S}$ (c); maximum atmospheric temperature $T_{\rm max}$ (d); maximum contribution from the molecular cooling to the total radiative cooling (e); maximum contribution from the Mlin term (f); maximum number density of H$_2$O at the lower boundary (g); maximum number density of CO (i); maximum contribution from $H_{\rm Hlin}$ term to the total heating (j); maximum number density of O$_2$ (l); maximum number density ratio of H(2P) to H(1S) (m); and maximum number density ratio of H(2S) to H(1S) (n). All parameters are shown against the atmosphere's mean particle weight and different model settings are shown in the legend.}
    \label{fig:main_out_gj9827d}
\end{figure*}

\subsection{Reduction of the atmospheric mass loss and bulk atmospheric properties}\label{sec:results_mass_loss_etc}%
\begin{figure*}
    \centering
    \includegraphics[width=0.8\linewidth]{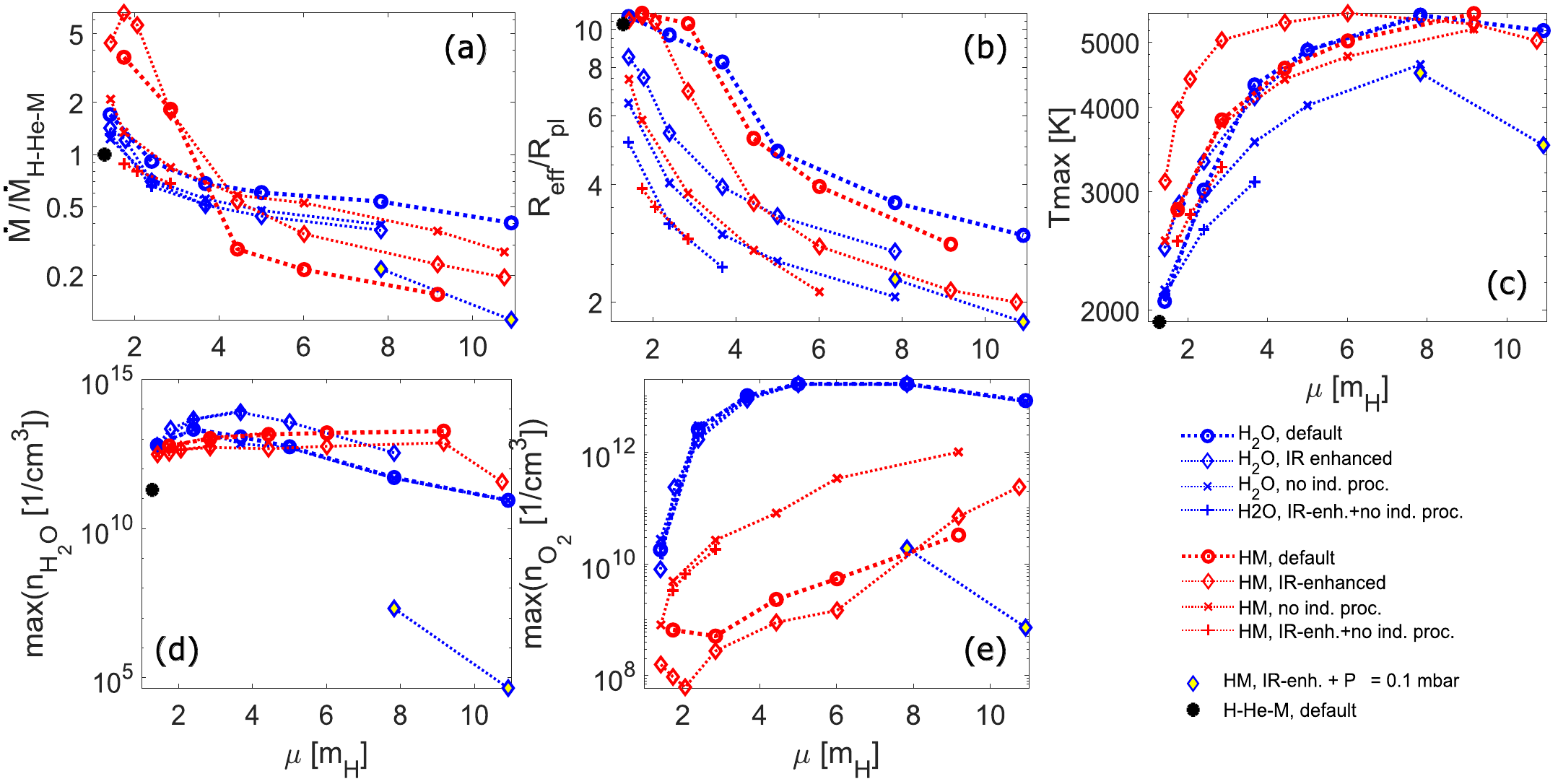}
    \caption{Summary of the simulation outputs for different model settings for TOI-238\,b: ratio of the mass loss rate to the default model (a); indicative effective radius of the EUV absorption \Reff\,(b); maximum atmospheric temperature $T_{\rm max}$ (c); maximum number density of H$_2$O at the lower boundary (d); maximum number density of O$_2$ (e). All parameters are shown against the atmosphere's mean particle weight and different model settings are shown in the legend.}
    \label{fig:main_out_toi238b}
\end{figure*}

%-
Figure\,\ref{fig:main_out_gj9827d} shows the dependence of some characteristic parameters on the mean weight of the atmospheric species for GJ\,9827\,d. The blue and red lines correspond to the water-enriched and metal-enriched atmospheres, respectively; shape of the markers indicates the model configuration: default model + default SED (circles), default model + IR-enhanced SED (diamonds), no induced processes + default SED (``x''), and no induced processes + IR-enhanced SED (``+''); finally, the colour filling in the markers, where present, indicates the special cases (reduced irradiation/alternative lower boundary pressure), as given in the legend. {We note that in figure\,\ref{fig:main_out_gj9827d} and anywhere further in the text wherever we quote any ratios between H-He and enriched atmospheres, these ratios are given for the cases with the same irradiation level/model settings.} The black circle marker shows the default case for the H-He atmospheres. Figure\,\ref{fig:main_out_toi238b} follows the same format and shows some of the characteristic parameters across the models performed for TOI-238\,b, where the results are considerably different to the GJ\,9827\,d.

For all model configurations, we see similar patterns. At relatively low atmospheric enrichment in heavy elements ($Z\sim10$), we see a small enhancement in the $\dot{M}$ relative to the H/He case, due to the extra heating at low altitudes becoming relevant while metal line cooling is still minor compared to the adiabatic term. With increasing enrichment, the mass loss ratio decreases down to $\sim0.2$ for the water-rich atmospheres and to $\sim 0.3$ for the metal-rich atmospheres (see $\dot{M}/\dot{M}_{\rm H-He-M}$ in panel a of Fig.\,\ref{fig:main_out_gj9827d}; note, that for each model configuration the ratio is calculated to the corresponding H-He model). The strongest reduction relative to the H-He atmospheres is seen for the default model configuration; however, the absolute values of the $\dot{M}$ (see Tab.\,\ref{tab:gj9827d_simlist0}) are smaller for the models without induced processes and with IR-enhanced SED, with the difference to the default case being larger for the H-He atmospheres and lower $\mu$ cases. Furthermore, the difference increases for lower irradiation cases. Thus, for the default case with $f^{\rm m}_{\rm H_2O} = 50\%$ the reduction is only a factor of $\sim0.8$; when moving the planet to 400\,K and 300\,K orbits, it reduces to $\sim0.6$ and $\sim0.065$, respectively (for the default case). Interestingly, the strongest reduction is seen not for the 300\,K-orbit case, but for the case with the XUV-level corresponding to the 400\,K orbit and the equilibrium temperature at the default value of $\sim610$\,K. It occurs, presumably, due to the more effective metal cooling at higher temperatures acting at higher altitudes than the molecular cooling (which is stronger at lower temperatures).

For TOI-238\,b, the atmospheric enrichment in metals becomes more effective than water-enrichment in reducing the atmospheric escape at $\mu\sim 4-5$ (see panel a of Fig.\,\ref{fig:main_out_toi238b}). The absolute values of both molecular ($C_{\rm molecule}$) and metal line cooling rates ($C_{\rm \Sigma M}$) are higher for TOI-238\,b; however, if the increase in molecular cooling is moderate (within a factor of a few), similar for the different atmospheric types, and fairly constant with $\mu$, the increase in the sum metal lines cooling rate (acting at high altitudes) is over an order of magnitude and significantly higher for the metal-enriched atmospheres. The absolute value of $C_{\rm \Sigma M}$ for metal-rich atmospheres increases steeply between $\mu\sim1-4$ ($\sim$ 200 times). The {absolute value of the metal heating follows the pattern similar to metal cooling (though it operates at different altitudes), but} the increase is only about 10 times over the same $\mu$ range; this leads to the stronger enhancement in the escape at low $\mu$ values for TOI-238\,b compared to the case of GJ\,9827\,d. {The relative contribution from metal heating to the total local heating rate, however, follows the same pattern as molecular cooling.}

Generally, the reduction of the escape occurs partly due to the increasing molecular/metal lines cooling rates and partly due to the increasing mean particle weight of the atmosphere, which makes the atmosphere significantly more compact. For the upper atmospheres, it can be seen well through the reduction of some characteristic distances, such as photodissociation and photoionisation fronts, exobase position, and the sonic point ($R_{\rm S}$). In Fig.\,\ref{fig:main_out_gj9827d}, we show the characteristic effective radius of the EUV absorption ($R_{\rm eff}$) and $R_{\rm S}$ in panels (b) and (c) for GJ\,9827\,d, respectively. $R_{\rm eff}$ is defined assuming the simplified model and only considering the hydrogen opacity and the photons with energy of 20\,eV \citep[as described in][]{kubyshkina2018grid}; it does not quite reflect the realistic picture of the photoionisation heating in the atmosphere %(photoionisation occurs over the wide range of altitudes and the peak in H$^+$ density is typically located at $\sim$1.1-2.0\,\Rpl) 
but is convenient to use as a proxy. One can notice, that for the low-irradiation cases, $R_{\rm eff}$ shifts closer to the planet (as the cooler atmospheres are more compact), while the sonic point remains at distances similar to the default cases. Both $R_{\rm eff}$ and $R_{\rm S}$ are similar between the water-rich and the metal-rich cases, implying that the differences in the atmospheric escape reduction do not come from the structural changes but rather from the differences in the radiative heating and cooling (namely, a higher contribution from the metal heating in the metal-rich case). For TOI-238\,b, the contraction of the atmosphere is counteracted by the higher equilibrium temperature, and the drop in the characteristic distances with $\mu$ is more shallow (panel b of Fig.\,\ref{fig:main_out_toi238b}); \Reff\ value remains roughly constant between $\mu\sim1-3$. We note that the increase in the XUV flux without an increase in \Teq\ does not affect the positions of the photoionisation/photodissociation fronts significantly.

Along with the reduction in the escape, the atmospheric contraction leads to the increase in the peak atmospheric temperatures (panel d of Fig.\,\ref{fig:main_out_gj9827d} and panel c of Fig.\ref{fig:main_out_toi238b}); this occurs due to the increase in density gradients and the narrowing of the photoionisation regions, similarly to the increase in the peak temperatures for more massive planets under the same conditions (thus, the typical peak temperatures for hot sub-Neptune-like and Jupiter-like planets are of the few thousand kelvin and $\sim10^4$\,K, respectively). For both planets, the increase in $T_{\rm max}$ slows down at $\mu\sim5-6$ and, for TOI-238\,b, the temperature slightly decreases again for $\mu$ above 7-8, as the increase in cooling becomes more relevant than the atmosphere's contraction.

The maximum contribution from the molecular cooling $C_{\rm molecule}$ to the total $C_{\rm rad}$ (panel e of Fig.\,\ref{fig:main_out_gj9827d}) remains constant with $\mu$ for the metal-rich atmospheres ($\simeq1$ near the lower boundary of the simulation domain), while it declines for the water-rich atmospheres; this decline, however, does not reflect any real physical process preventing the molecule formation but the shift of the photodissociation fronts to the lower altitudes, as discussed in Sec.\,\ref{sec:results_heating_and_cooling} (as, in particular, seen for the higher lower boundary pressure case near $\mu=4$). For the more appropriate boundary conditions, we expect the behaviour of $C_{\rm molecule}/C_{\rm tot}$ for the water-rich atmospheres to be similar to the metal-rich case. The absolute values of $\max(C_{\rm molecule})$, for the metal-rich atmospheres, increase with $\mu$ by the factor of 2-5, depending on the model configuration (stronger for the models without the induced processes).

On the contrary, the maximum contribution from the sum of metal line cooling processes $C_{\Sigma M}$ (panel f of Fig.\,\ref{fig:main_out_gj9827d}) increases with $\mu$ for both atmospheric types (as it is achieved typically at much higher altitudes near $\sim10^{-2}$\,\dyncm). For a given $\mu$, this contribution is higher for the metal-rich atmospheres, where it becomes a dominant cooling agent at $\mu\simeq3$. The absolute values of $\max(C_{\Sigma M})$ are roughly constant with $\mu$ for the water-rich atmospheres and increase by the factor of 4-50 for the metal-rich atmospheres (this value is significantly smaller for the IR-enhanced SED models at high $\mu$, while similar at low $\mu$; however, the relative contribution is similar for all model configurations).

The maximum contribution from the Mlin heating term to the total heating is 100\% at the lower boundary for all models with the induced processes (and is zero without). For the Hlin term (panel j of Fig.\,\ref{fig:main_out_gj9827d}), it becomes a dominant heating process in the region below the one dominated by the classic H photoionisation heating for the water-rich atmospheres and the models with induced processes; without the induced processes, it still can contribute up to 20-30\% in this region. For the water-rich atmospheres, the maximum contribution from the Hlin term is roughly constant with $\mu$, but for the metal-rich atmospheres it decreases significantly at high $\mu$.

We further show the maximum number densities of H$_2$O, CO, and O$_2$ molecules for GJ\,9827\,d (panels g, i, and l of Fig.\,\ref{fig:main_out_gj9827d}, respectively). For the water-rich atmospheres, these values are strongly affected by the lower boundary issue discussed above, so we focus on the metal-rich atmospheres here. The number densities of most of the molecules maximize at the lower boundary of the simulation domain, except for some mono-molecules, such as O$_2$, which density in some cases maximizes near the H$_2$O photodissociation boundary. The molecule abundances do not increase indefinitely with the increasing heavy elements fraction; instead, the abundances of H$_2$O and CO increase by about 10 times from the H-He case to $Z=10Z_{\odot}$ and by another 10 times between $\mu\sim1.2-3$; at higher $\mu$, maximum abundances of both reach a saturation at $10^{13}$\,${\rm cm^{-3}}$. Between the two molecules, CO is not only more stable at high altitudes (see Fig.\,\ref{fig:molecules_HM}) but also is weakly sensitive to the changes in $\mu$ and the model assumptions. For TOI-238\,b, behaviour of CO is similar to the GJ\,9827\d case, while for H$_2$O, most of the increase in the number density occurs over the narrower $\mu$ range of 1.2-1.4 (see panel d of Fig.\,\ref{fig:main_out_toi238b}). The maximum abundance of O$_2$ follows the pattern similar to CO for the water-rich atmospheres for both planets (steep increase below $\mu\sim3$ and roughly constant above; see panel l of Fig.\,\ref{fig:main_out_gj9827d} and panel e of Fig.\,\ref{fig:main_out_toi238b}). For the metal-rich atmospheres, its abundance grows more steadily between $\mu\sim1-10$. Most of the molecular abundances are fairly insensitive of the model configuration, except that H$_2$O is slightly more abundant for the IR-enhanced SED in the water-rich atmospheres and the O$_2$ abundance in the metal-rich atmospheres at the hotter TOI-238\,b depends on the inclusion of the induced processes (as they affect the heating at low altitudes and, hence, the altitude  and conditions where $n_{\rm O_2}$ peaks).

Finally, panels (m) and (n) of Fig.\,\ref{fig:main_out_gj9827d} show the variations in the peak abundances of hydrogen in 2P and 2S excited states, respectively. One can see that the number of atoms in 2S state increases with $\mu$ for all models (with saturation/peak near $\mu\sim 6$), while the number of atoms in 2P state remains roughly constant (for the water-rich atmospheres) or slowly declines (metal-rich atmospheres). The contribution of the Hlin process to the total heating correlates with the number of atoms in 2P state (see panels j and m of Fig.\,\ref{fig:main_out_gj9827d}).

\section{Discussion}\label{sec:discuss}
%-
\subsection{Influence of the model assumptions}\label{sec:discuss_model_ass}%
%-
The CHAIN model relies on a few assumptions which might affect our predictions, such as a one-fluid approximation for the hydrodynamic model (meaning that the atmospheric dynamics is fully controlled by the hydrogen flow and all other elements are passively dragged along), no condensation of the atmospheric species, fixed boundary conditions, and the limited chemical framework. Concerning the latter, one of the main limitations of the Cloudy chemical solver is the absence of some relevant molecules, in particular CO$_2$, which is known to be an effective coolant \citep[e.g.][]{Yoshida2024PEPS...11...59Y} and is expected to be present in the similar domain as H$_2$O molecules. The newest release of Cloudy \citep[][]{Gunasekera2025arXiv250801102G} includes an extended chemical framework, but still no CO$_2$. This, however, can have a limited effect on our results% (though can hardly be tested at the present state)
, as the above discussion indicates that the heating and cooling processes of atomic metals have a stronger influence on our results, as they occur at higher altitudes. These processes are mainly affected by the assumptions discussed in more detail below.

\subsubsection{Condensation of the atmospheric species}\label{sec:discuss_condensation}

The condensation of the atmospheric species can lead to the absence of the specific elements, included in our models, in the upper atmospheres, which can affect both heating and cooling processes. To assess the possible effects of condensation, we adopted the condensation temperatures given in \citet{Wood2019AmMin.104..844W} for the {100\,\dyncm} pressure, which is fairly close to our lower boundary pressures. These temperatures are given for the 50\% of the species condensed; however, for the rough test we assume that if the temperatures at the lower boundary in our models are lower than these values, the specific elements were condensed (and removed from the upper atmosphere through the rain-out) at higher pressures. We therefore remove these species (which is, most of the heavy elements for GJ\,9827\,d, except the C-N-O group, sulphur, and a few others having no considerable effect, and heavier metals Mg-Si, Ca-V, and Fe-Ni for TOI-238\,b) from our chemical framework and re-run the last iteration of Cloudy model (for the converged bulk atmosphere parameters) under this new assumption. We note that this test only allows to estimate the scale of the effect expected from condensation. Below, we show results for the cooler planet, GJ\,9827\,d, where condensation is more relevant; for TOI-238\,b, results are qualitatively similar but changes relative to the default models are much smaller.

\begin{figure}
    \centering
    \includegraphics[width=0.6\linewidth]{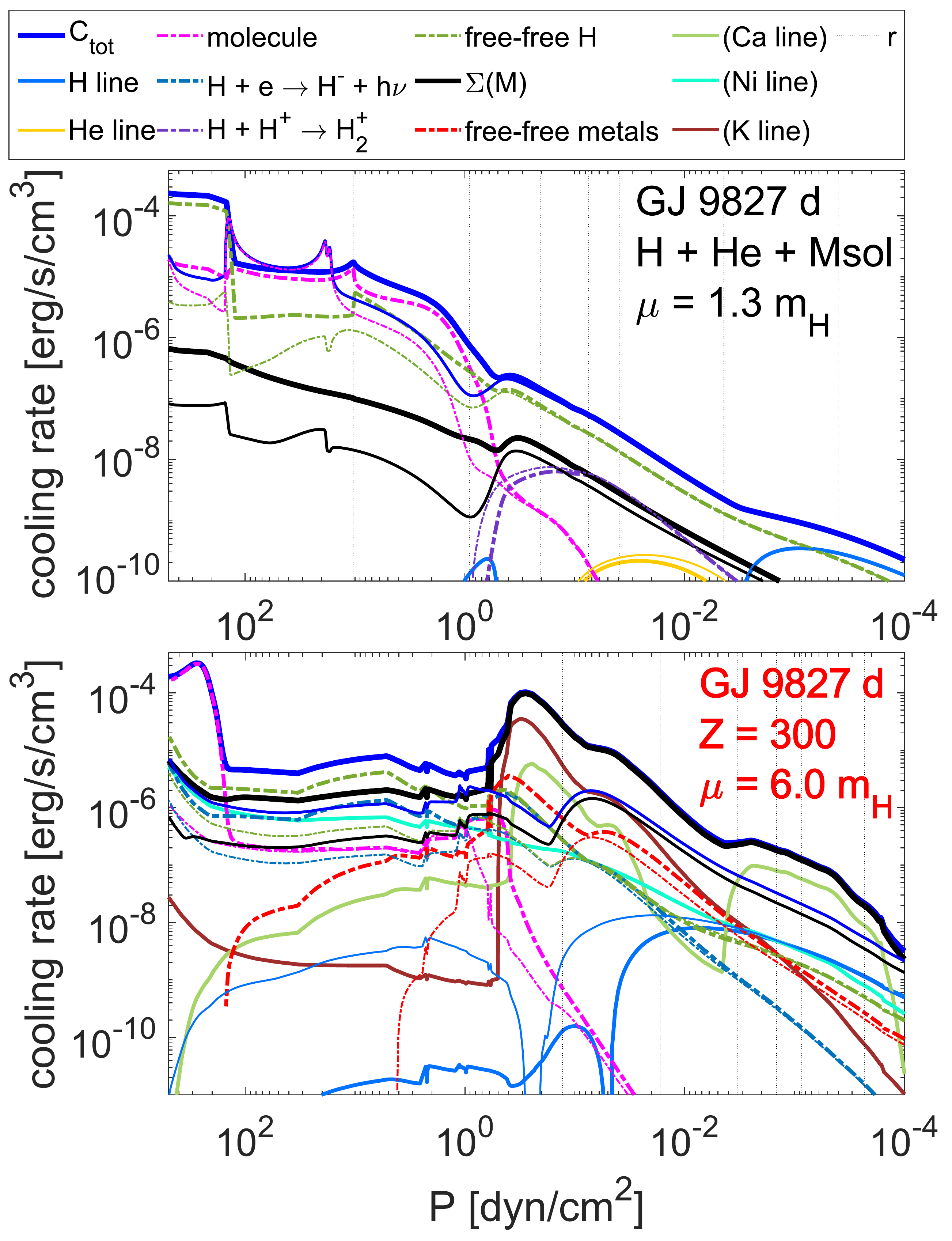}
    \caption{Radiative cooling processes assuming the default abundances (thicker line) and accounting for the condensation (thinner lines) for GJ\,9827\,d and the H-He (top) and $Z=300Z_{\odot}$ (bottom) cases. The line types are explained in the legend. The metal line processes that are missing when accounting for the condensation are shown in the brackets. The lines defined as r denote the radial distances of 1.1, 1.2, 1.3, 1.4, 1.5, 2.0, and 3.0\,\Rpl.}
    \label{fig:cool_condensation}
\end{figure}
\begin{figure}
    \centering
    \includegraphics[width=0.6\linewidth]{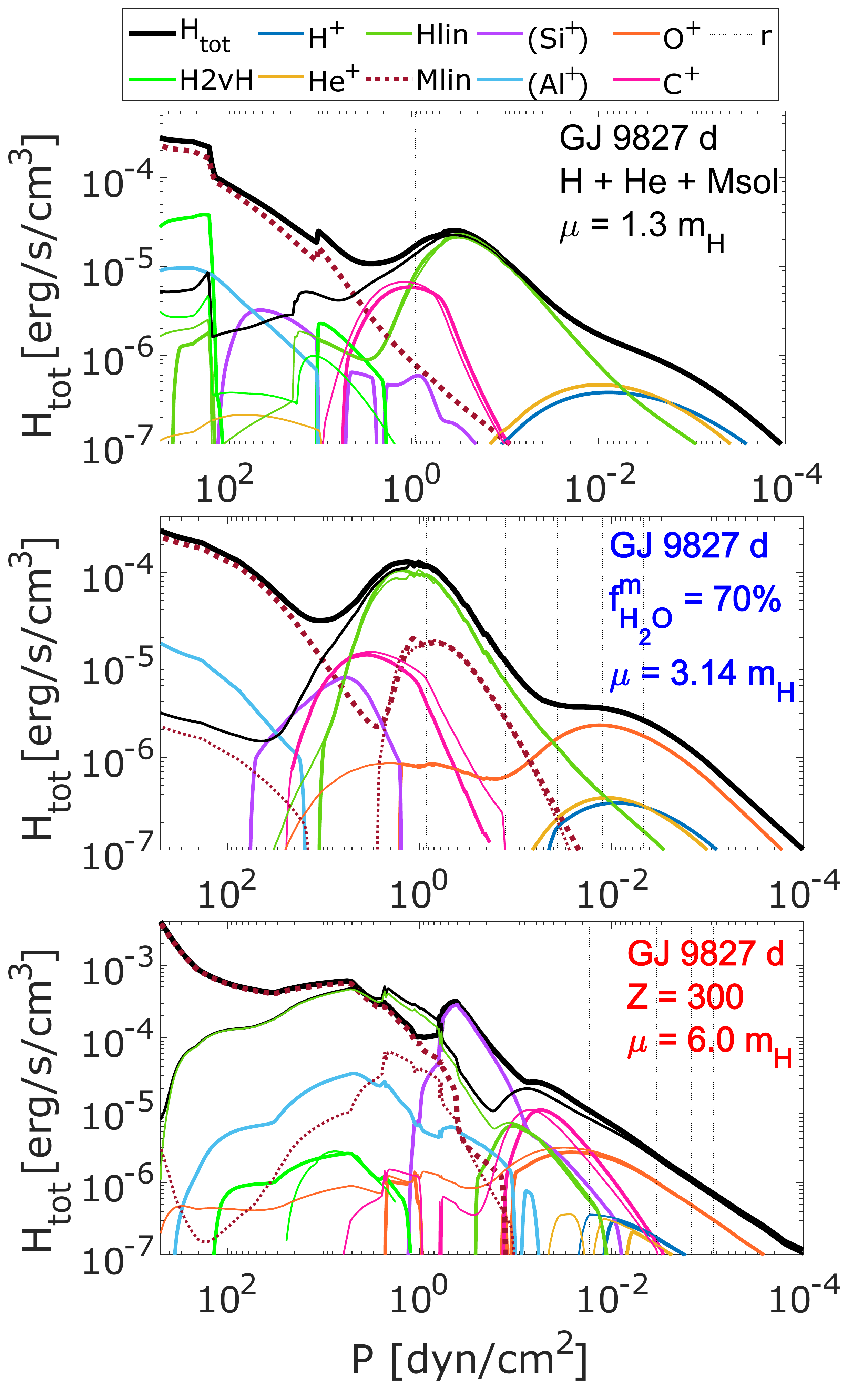}
    \caption{Volume heating rates assuming the default abundances (thicker line) and accounting for the condensation (thinner lines) for GJ\,9827\,d and the H-He (top), $f^{\rm m}_{\rm H_2O} = 70\%$ (middle), and $Z=300Z_{\odot}$ (bottom) cases. The line types are explained in the legend. The heavy element ionisation processes that are missing when accounting for the condensation are shown in the brackets. H2vH term denotes the collisional de-excitation of the vibrationally excited H$_2$. The lines defined as r denote the radial distances of 1.1, 1.2, 1.3, 1.4, 1.5, 2.0, and 3.0\,\Rpl.}
    \label{fig:heat_condansation}
\end{figure}
%
%\begin{figure}
%    \centering
%    \includegraphics[width=0.6\linewidth]{Hpops_pp_comp_cond.pdf}
%    \caption{Populations of some H species and heavy ions assuming the default abundances (thicker line) and accounting for the condensation (thinner lines) for GJ\,9827\,d and the H-He (top) and $Z=300Z_{\odot}$ (bottom) cases.}
%    \label{fig:hpops_condensation}
%\end{figure}

In Fig.\,\ref{fig:cool_condensation}, we compare the cooling processes in our default models (thicker lines) with the model accounting for the condensation (thinner lines) for the H-He atmosphere and the model with $Z = 300 Z_{\odot}$. Accounting for the condensation leads to the decrease in cooling in a few zones; most prominent are (i) the decrease in the molecular cooling at pressures above 100\,\dyncm (for all atmospheric types) due to a decrease in the molecule's variety, and (ii) the decrease in the metal line cooling %at pressures above $\sim1$\,\dyncm for water-rich atmospheres (which has weak effect on the total cooling for most altitudes) and 
at pressures above $\sim10^{-3}$\,\dyncm for the metal-rich atmospheres (which leads to a significant reduction in the total cooling at high altitudes).
Fig.\,\ref{fig:heat_condansation}, in turn, compares the heating processes with and without condensation for the same cases as considered in Fig.\,\ref{fig:heat_compn_noind}. One can see, that accounting for the condensation leads to the decrease in the heating at pressures above $\sim1$\,\dyncm in all cases, and above $\sim0.1$\,\dyncm for the metal-rich atmosphere. Generally, the changes to the cooling and heating processes are similar to the changes corresponding to the exclusion of the induced processes; therefore, we expect that the changes to the bulk parameters in self-consistent simulations should also follow similar patterns.

\subsubsection{One-fluid approximation}\label{sec:discuss_one-fluid}

The ability of the hydrogen outflow to drag the heavier species along is strongly limited by the strength of the outflow, the local atmospheric densities and temperatures, and the collisional cross-sections between different species (hence also the ionisation state of the atmosphere). The proper assessment of this effect requires employing a multi-fluid hydrodynamic approximation \citep[e.g.][]{erkaev2023MNRAS.518.3703E,Schulik2023MNRAS.523..286S}, which is, however, challenging in combination with a complex chemical framework. In case of one-fluid models, the drag can be assessed in post-processing, for example, using the formulation presented in \citet{Catling2017aeil.book.....C}, which is commonly employed in atmospheric escape studies \citep[e.g.][]{Linssen2024A&A...688A..43L,Louca2025arXiv250506013L}. It is limited to defining the critical mass ($\mu_{\rm crit}$), which is the maximum mass of the element that can be dragged along with atmospheric hydrogen, at the photosphere.

For the models considered above \citep[following the approach described in][]{Linssen2024A&A...688A..43L}, the value of $\mu_{\rm crit}$ at the photosphere is about 6-20 %(hydrogen masses) for the metal-enriched atmospheres and between 10 and 100 (with the minimum reached near $\mu\sim 4$) for the water-rich atmospheres 
in the case of GJ\,9827\,d; for TOI-238\,b, $\mu_{\rm crit}$ %is similar for the water-rich case and 
changes between $\sim$10-100. % for the metal-rich atmospheres. 
This suggests that a significant fraction of the heavy elements would be dragged along by hydrogen outflow and escape from the planet. 
However, we note that these values depend strongly on the (fixed) lower boundary conditions. Furthermore,  for non-isothermal atmospheres $\mu_{\rm crit}$ is not necessarily constant with altitude. Using the same formulation, we find that the value of $\mu_{\rm crit}$ decreases with pressure at low altitudes. It reaches the minimum value (1-1.4 for GJ\,9827\,d and 1.3-3.5 for TOI-238\,b, for the models considered here) at intermediate pressures (10$^{-3}$-10$^{0}$\,\dyncm ) and increases slightly again at high altitudes due to the increasing ion fraction. In this case, one would expect that only hydrogen can escape. This estimates, however, have to be verified in a self-consistent simulation, which we leave for the future studies.

%\subsection{Comparison with other models}\label{sec:discuss_compare}%
%
\subsection{Implications for planetary evolution}\label{sec:discuss_evolution}%
The results presented above have direct implication for the atmospheric stability in the evolution context. Besides the evident improvement of the atmospheric stability for hot sub-Neptunes in the case of strong atmospheric enrichment with water or other heavy elements, one has to account for the following two points. First, the atmospheric composition is not constant in time: as discussed above, the light elements are expected to escape at higher rates than the heavy ones. The tests similar to the one discussed in Sec.\,\ref{sec:discuss_one-fluid} performed on the grid of models from \citet{kubyshkina_fossati2021} suggest that the heavy elements are dragged along with hydrogen for highly irradiated and inflated planets \citep[which corresponds typically to a very short early period of planet's evolution, e.g.][]{kubyshkina2022AN....34310077K}, but this regime transitions quickly into the regime where only H atoms escape as the planet cools and contracts, with the intermediate phase of differential escape existing for a relatively narrow range of parameters. Therefore, one can expect that the atmospheric metallicity increases in time, and so does the reduction in the escape (relative to H-He atmospheres) for given parameters.

Second, as we have shown above, an increase in the metallicity leads to atmospheric contraction. However, what was not considered here is that it concerns not only the characteristic distances in the upper atmosphere, but also the lower atmosphere and, in particular, the position of planet's photosphere (\Rpl). Thus, for the similar mass fraction of the atmosphere and planetary core parameters, the H-He atmosphere is more expanded than the water-rich or metal-rich atmospheres; hence, the interaction area with stellar XUV is also larger for H-He atmospheres, which aggravates the differences in the escape between the H-He and the enriched atmospheres. Therefore, the actual reduction in the escape with increasing metallicity for a given atmospheric mass (and not a given radius, as in the present study) is expected to be higher than discussed above.

\section{Conclusions}
\label{sec:conclusions}%
We analysed the differences in dominant heating and cooling processes and bulk atmospheric outflow properties under different assumptions on atmospheric composition for the two highly irradiated compact sub-Neptune-like planets: more moderate GJ\,9827\,d (where water presence in the atmosphere was confirmed by JWST observations) and more extreme TOI-238\,b (proposed to be a possible hot water-world planet). We considered water-rich atmospheres with water mass fractions up to 90\% and the metal-rich atmospheres with metal abundances up to 500$\times$solar. We found that

\begin{itemize}
    \item For the both atmospheric types, the $C_{\rm rad}$ increases with atmospheric metallicities.
    \item The effectiveness of the molecular cooling processes at pressures above $\sim1$\,\dyncm is limited by the molecules' stability in the high-irradiation environment: we see a saturation in the molecular abundance with increasing  mean atmospheric particle mass, with the saturation threshold at $\mu\sim$\,2-4. %, depending on the specific planet/model configuration. 
    %Furthermore, the molecular cooling is limited to the atmospheric regions with pressures above $\sim1$\,\dyncm.
    \item Metal line cooling processes are effective contributors to the increase in the total cooling of the atmosphere. Along with the increase in the cooling, however, we see an increase in the local heating rates, which counteracts the decrease in atmospheric escape rates with metallicity. 
    \item We find a large contribution to the heating from the induced processes, which was not found for more moderate planets. These processes affect the absolute values of the $\dot{M}$ (up to a factor of 3, with a strongest effect seen in the atmospheres with low metallicity), but do not affect qualitatively the dependence of the bulk atmospheric properties on metallicity.
    \item For all models, we see a decrease in the $\dot{M}$ with increasing $\mu$ (down to factor of 0.25 for GJ\,9827\,d, 0.045 for the low {XUV} radiation case, and to factor of 0.15 for TOI-238\,b), except for a narrow region of low enrichment levels ($\mu<2$ for the H$_2$O-rich and $\mu<3$ for the metal-rich atmospheres), where the escape is enhanced relative to a pure H-He composition due to the additional metal heating. 
    \item We find that the reduction in the escape at a given $\mu$ depends strongly on the irradiation conditions at the planet (for GJ\,9827\,d). The largest decrease among the considered models corresponds to the reduced XUV at constant \Teq\ (bolometric heating).
\end{itemize}
 
%It is accompanied by the atmospheric contraction and shift of the characteristic atmospheric distances closer to the planet. 
For the convenience, we summarise the key outputs in Tab.\,\ref{tab:summary}. In general, our results suggest that the reduction in the atmospheric escape with increasing water/metal enrichment for highly irradiated planets can be smaller that expected for more moderate conditions. However, this reduction, along with the enrichment increasing over time due to the fractionation, can be sufficient for the earlier cessation of the hydrodynamic blow-off and for atmospheres to be stable even under harsh conditions.

\begin{acknowledgements}
D.K. was supported by a Schr\"odinger Fellowship supported by the Austrian Science Fund (FWF) project number J4792 (FEPLowS). %We thank the anonymous referee for their comments that helped improving the presentation of the results. 
J.A.E. acknowledges support through the European Space Agency (ESA) Research Fellowship Programme in Space Science. This work has been carried out within the framework of the NCCR PlanetS supported by the Swiss National Science Foundation under grants 51NF40\_182901 and 51NF40\_205606.
C.P.-G. acknowledges support from the E. Margaret Burbidge Prize Postdoctoral Fellowship from the Brinson Foundation as well as funding received through the Suzuki fellowship.
\end{acknowledgements}

%\section*{Data Availability}

%The data and models underlying this study are available at a reasonable request to the corresponding author. 


\bibliographystyle{aa} % style aa.bst
\bibliography{HeavyMetal} % your references Yourfile.bib

\begin{appendix}

\section{Additional information}\label{apx:summary_table}
\begin{figure}
    \centering
    \includegraphics[width=\linewidth]{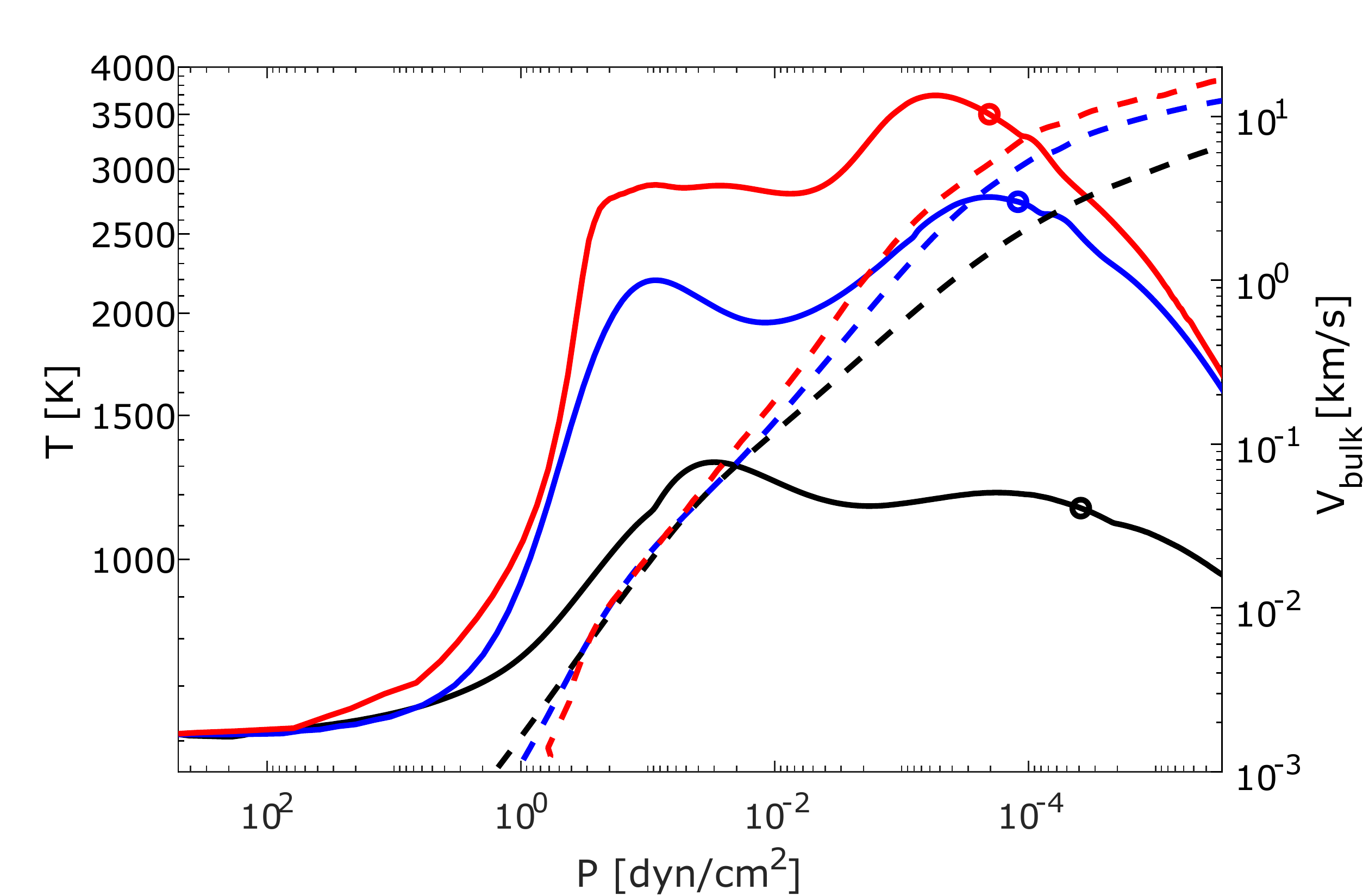}
    \caption{Temperature (solid lines) and bulk velocity (dashed lines) corresponding to the ``default'' model cases considered in Figures\,\ref{fig:cool_compn_noind}--\ref{fig:heat_compn_noind}.}
    \label{fig:TV_gj9827d}
\end{figure}

In Table\,\ref{tab:summary}, we summarise our main findings for GJ\,9827\,d and TOI-238\,b. Specifically, we list the species and pressure intervals relevant for the radiative heating and cooling processes, most abundant molecules, and the escape reduction relative to the case of H-He-dominated atmospheres. The results are given for the default model configuration and the default SED, unless specified otherwise; in other cases, the results are qualitatively similar. Figure\,\ref{fig:TV_gj9827d} shows the temperature and bulk velocity profiles discussed in Sec.\,\ref{sec:results_heating_and_cooling}.
\begin{table*}[th]
    \centering
    \caption{Summary of the key results. For the heating, the relevant species are given in the format with/without induced processes. For the cooling, ``molecules'' refers to the heavy (non-hydrogen bearing) molecules. {For the convenience of the reader, we give the pressures in mbar.}}
    \begin{tabular}{c|c|c|c}
    \toprule
     & H-He-dominated & Water-rich & Metal-rich \\
     \midrule
      heating: &  & &\\
      relevant species & H$^+$, H(2P)/ H$^+$ & H(2P), O$^+$/H$^+$, O$^+$ & Al, Mg, Ca, \& Fe \\
      relevant pressures & $10^{-7}-10^{-3}$\,mbar & $10^{-5}-10^{-2}$\,mbar & $\gtrsim10^{-4}$\,mbar \\
      \hline
      cooling: &  & &\\
      relevant species & H + molecules & H + molecules & K, Ca, Ni, Mg, Na \\
      relevant pressures & $\gtrsim10^{-3}$\,mbar & $\gtrsim10^{-3}$\,mbar & $\gtrsim5\times10^{-6}$\,mbar \\
      \hline
      most abundant molecules: &  &  &  \\
      $\gtrsim10^{-3}-10^{-1}$\,mbar* & H$_2$ & H$_2$, H$_2$O & H$_2$\\
      $10^{-4}-10^{-2}$\,mbar & CO & CO & CO \\
      * & $n_{\rm H_2O}\simeq n_{\rm CO}$ & $n_{\rm H_2O} \gg n_{\rm CO}$ & $n_{\rm H_2O}\simeq n_{\rm CO}$\\
      \hline
      escape reduction & & & \\
       at $\mu=4$\,$m_{\rm H}$ & -- & 0.4 (GJ)/0.5 (TOI) & 0.6 (GJ)/0.5 (TOI) \\
       at $\mu=8$\,$m_{\rm H}$ & -- & 0.25 (GJ)/0.4 (TOI) & 0.4 (GJ)/0.2 (TOI) \\
      \bottomrule
    \end{tabular}
    \label{tab:summary}
\end{table*}
%blahblah

\section{Metal-free atmospheres}\label{apx:results_met-free}
To evaluate the general effect of metals in H-He-dominated atmospheres for the test planets, we ran the models also for H-He atmospheres without metals and pure H atmospheres. We found that in these cases, both atmospheric heating and cooling drop by about an order of magnitude at pressures above $\sim0.1$\,\dyncm; both heating and cooling are dominated by hydrogen-driven processes. For cooling, the free-free interactions of hydrogen dominate the cooling at lowermost altitudes, recombination of H$^+_2$ dominates near the photodissociation front, and H-line cooling dominates at high altitudes. The heating is dominated by the Hlin term and photoionisation of H; the latter increases significantly compared to the case of H-He atmospheres with solar metal abundance, and the former controls the heating at low altitudes. The total mass loss rates increase from H-He atmosphere with solar metal abundances to pure H atmospheres by a factor of $\sim1.6$ independently of SED; for H-He atmospheres with and without metals, the escape rates are about the same (which implies that most of the change is due to the change in $\mu$). 

\section{Adiabatic wind advection}\label{apx:results_wind-adv}
\begin{figure}[b]
    \centering
    \includegraphics[width=0.6\linewidth]{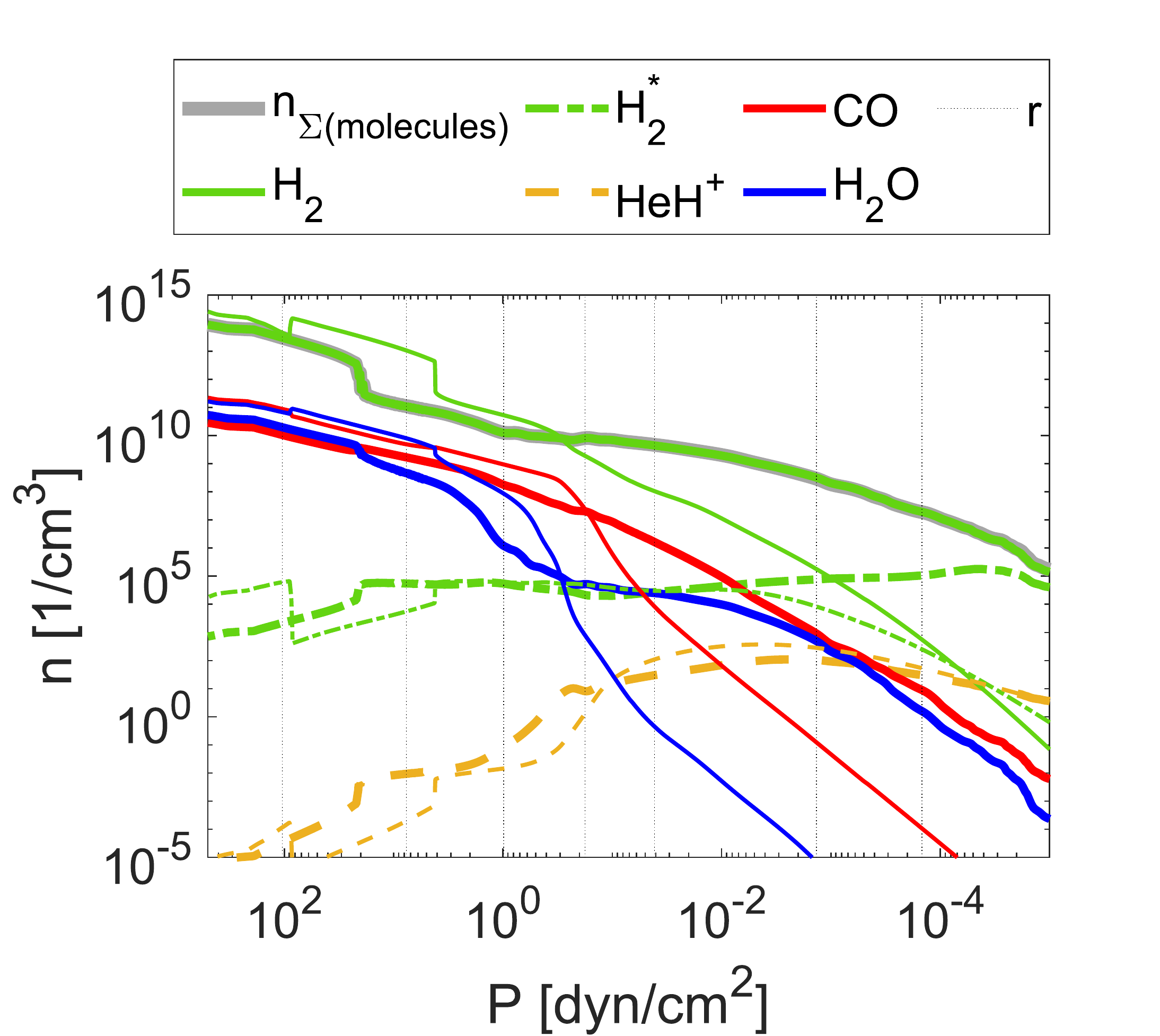}
    \caption{Molecular abundances with (thicker lines) and without (thinner lines) accounting for the adiabatic wind in the advection term in Cloudy. Line types are explained in the legend. The lines defined as r denote the radial distances of 1.1, 1.2, 1.3, 1.4, 1.5, 2.0, and 3.0\,\Rpl.}
    \label{fig:molecules_wind}
\end{figure}
The default configuration of Cloudy code is hydrostatic, implying that the chemical solver does not account for the hydrodynamic flow when performing the advection calculations; however, such a correction to the element advection in Cloudy can be prescribed externally (see details in K24). As this addition is computationally expensive, we do not include it in the default model configuration. We did, however, test the effect of the wind advection for the H-He atmosphere case of GJ\,9827\,d. As in K24, we did not see any significant changes to the mass loss rate and found that with the wind advection the atmospheric temperatures are slightly lower at low altitudes and higher near and above the sonic point (where the atmosphere becomes close to isothermal in this case), than otherwise. Including the wind advection further leads to the shift of the photoionisation and photodissociation fronts away from the planet (i.e., to the decrease in ion abundances and increase in molecular abundances at high altitudes; see Fig.\,\ref{fig:molecules_wind}). Though likely relevant for the interpretation of the observations, these changes do not affect significantly any bulk properties of the outflow; therefore, we do not expect them to be relevant for the present study.

\section{Lower boundary pressure}\label{apx:results_pressure}
For both planets considered here, the atmospheric escape occurs in the XUV-driven regime; therefore, the lower boundary conditions are not expected to have a strong influence on the results \citep[if vary in reasonable ranges; e.g.][]{Reza2025A&A...694A..88R}, though most of the tests were performed for H-He atmospheres without enrichment. We tested the lower boundary pressure of $10\times$ the default value for the $f^{\rm m}_{\rm H_2O} = 70\%$ for GJ\,9827\,d (for the region where total number density is above $10^{15}$\,cm$^{-3}$, we extend the cloudy predictions for heating/cooling rates at constant value; see the discussion in K24) and of $0.1\times$ the default value for the $f^{\rm m}_{\rm H_2O} = 90\%$ for TOI-238\,b (for both cases, we used the version with the IR-enhanced SED). For the higher pressure case, we did not see any significant changes in the bulk properties of the outflow; the atmospheric temperature increases slightly and the mass loss rate increases by a factor of $\sim1.6$ due to a slightly higher outflow velocity (see the blue diamonds markers with yellow face colour in Fig.\,\ref{fig:main_out_gj9827d}). There are no qualitative changes to either heating or cooling processes, except for a narrow peak in the molecular cooling at $\sim5000$\,\dyncm\ not accounted for at the default lower boundary pressure (of 1000\,\dyncm), which corresponds to the recombination of H$_2$ and H$_2$O (for both molecules).

In case of the boundary pressure $0.1\times$ of the default value, the predicted temperature of the outflow decreases in the whole modelling domain, the difference reaching up to 20\% near the temperature maximum; the bulk velocity of the outflow also decreases, but the total mass loss rate only decreases by the factor of $\sim1.7$. In this case, the heating rate at the lower boundary decreases by about a factor of 5 if comparing to the same pressures in the default case; the same occurs for total molecular abundances, though the relative contribution from different molecules remains the same. Both effects are local.

\section{Heating and cooling for TOI-238\,b}
\label{apx:extras}
\begin{figure*}
    \centering
    \includegraphics[width=0.33\linewidth]{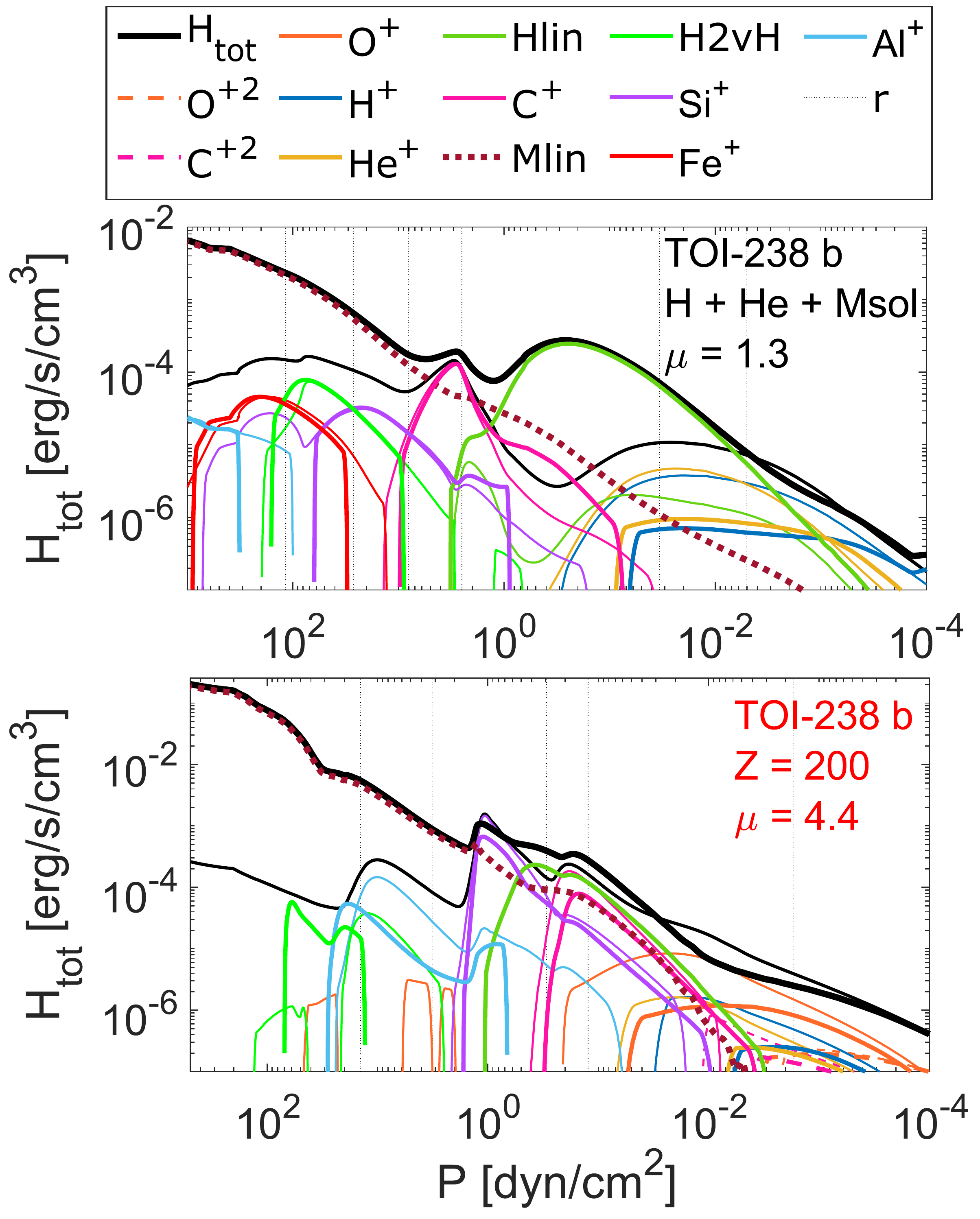} \includegraphics[width=0.33\linewidth]{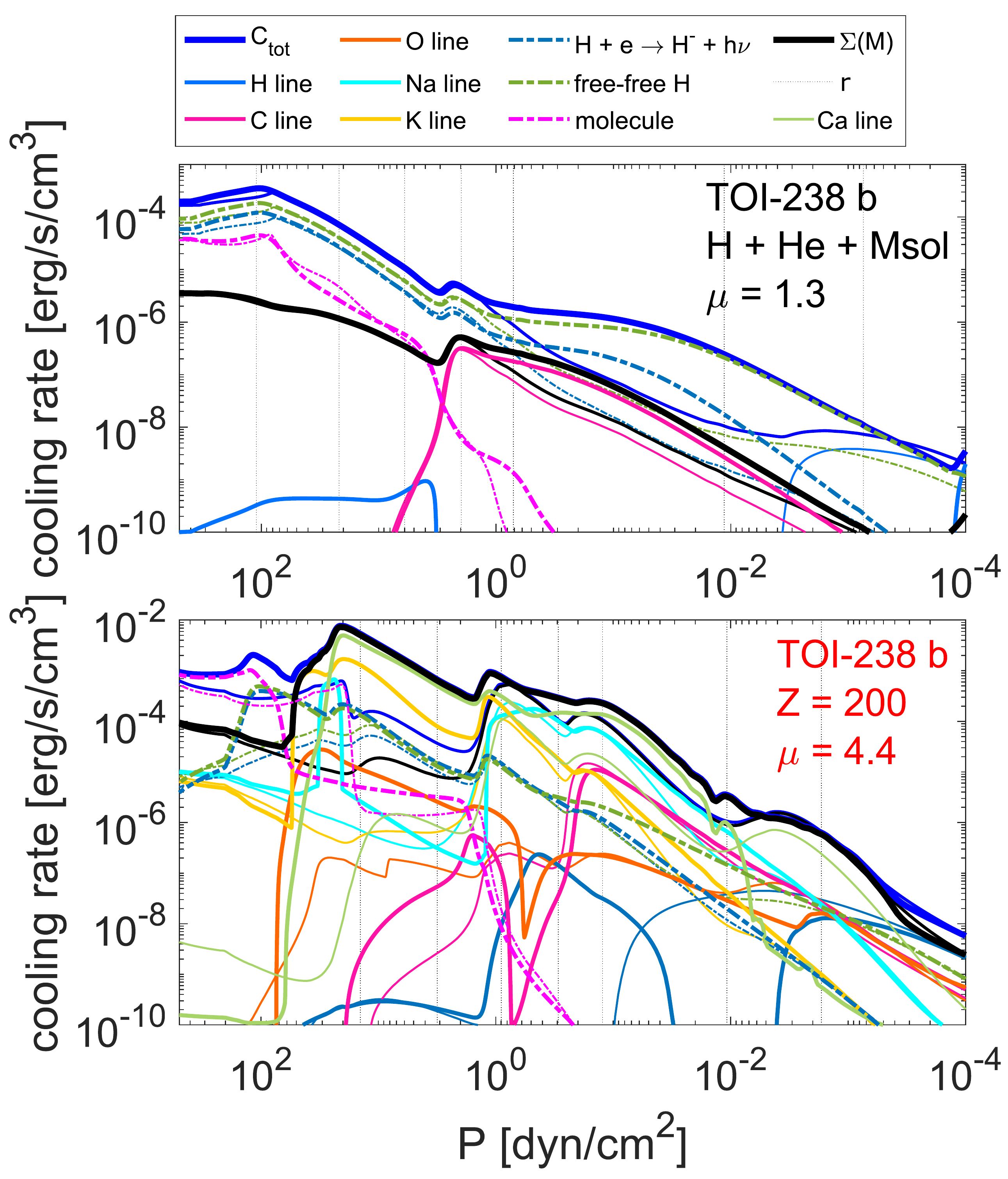} \includegraphics[width=0.25\linewidth]{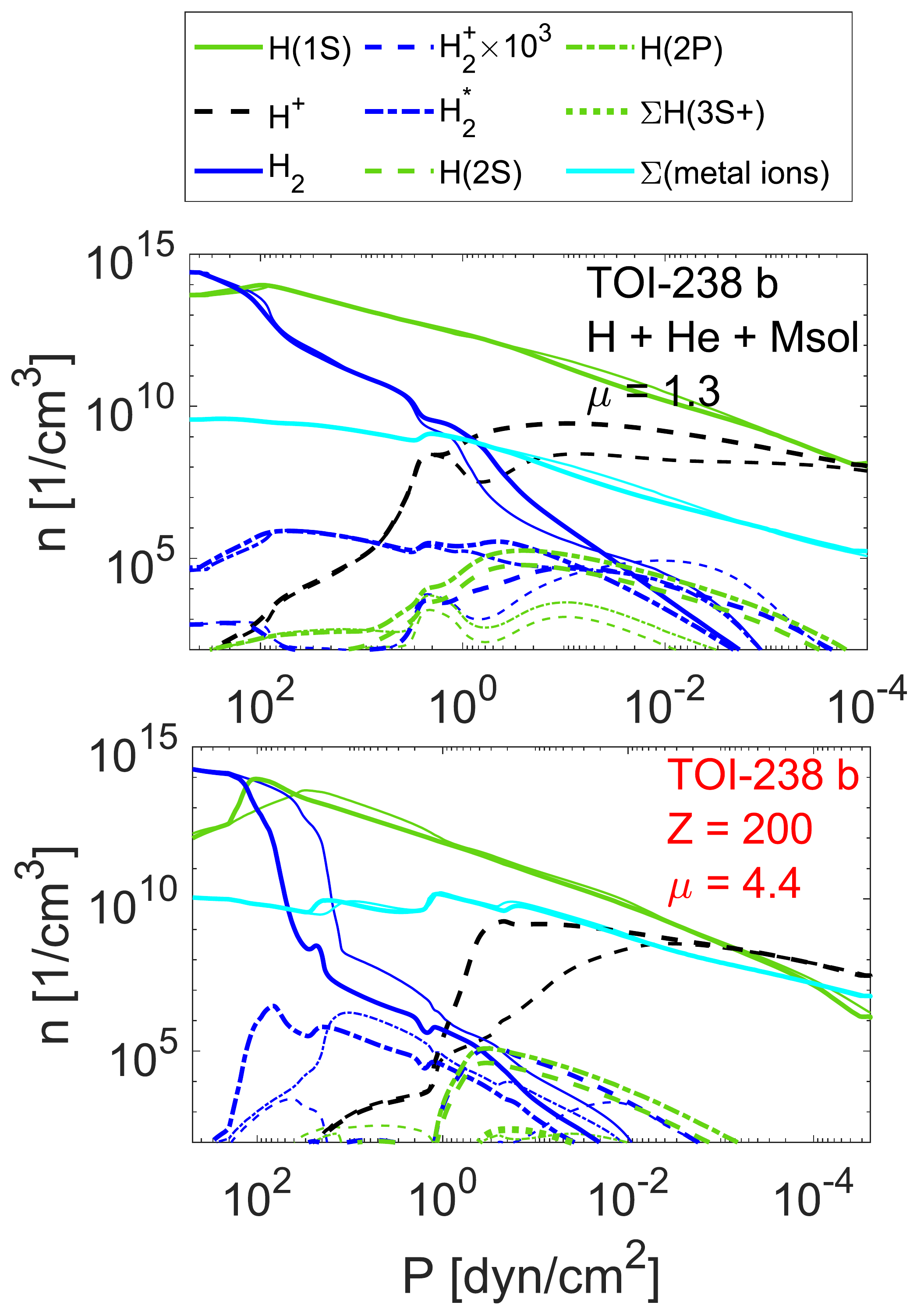}
    \caption{Left: Radiative cooling rates for TOI-238\,b: the H-He atmosphere with solar metallicity (top panel) and the atmosphere enriched in metals 200 times over solar metallicity (bottom panel). The plot format matches that of Fig.\,\ref{fig:cool_compn_noind}. Center: Volume heating rates for TOI-238\,b: the H-He atmosphere with solar metallicity (top panel) and the atmosphere enriched in metals 200 times over solar metallicity (bottom panel). The plot format matches that of Fig.\,\ref{fig:heat_compn_noind}. Right: Population of hydrogen species and heavy (metal) ions: the H-He atmosphere with solar metallicity (top panel) and the atmosphere enriched in metals 200 times over solar metallicity (bottom panel). The plot format matches that of Fig.\,\ref{fig:hpops_compn_noind}.}
    \label{fig:cool-heat-pop_toi238b}
\end{figure*}

%
%In Figures\,\ref{fig:cool_toi238b}, \ref{fig:heat_toi238b}, and \ref{fig:hpops_toi238b}, we show the radiative cooling, radiative heating, and populations of some hydrogen species for TOI-238\,b, respectively. The format of the figures matches that of Fig.\,\ref{fig:cool_compn_noind}--\ref{fig:hpops_compn_noind}. 
In Figure\,\ref{fig:cool-heat-pop_toi238b}, we show the radiative cooling (left), radiative heating (middle), and populations of some hydrogen species (right) for TOI-238\,b. The format of the plots matches that of Fig.\,\ref{fig:cool_compn_noind}--\ref{fig:hpops_compn_noind}.
As the results for the water-rich atmospheres are qualitatively similar to the H-He case, we omit them here and only show the H-He case and the case of the metal enrichment of $Z = 200$.
%
%\begin{figure}
%    \centering
%    \includegraphics[width=0.6\linewidth]{cool_pp_comp_noind_toi238b_upd.pdf}
%    \caption{Radiative cooling rates for TOI-238\,b: the H-He atmosphere with solar metallicity (top panel) and the atmosphere enriched in metals 200 times over solar metallicity (bottom panel). The plot format matches that of Fig.\,\ref{fig:cool_compn_noind}.}
%    \label{fig:cool_toi238b}
%\end{figure}
%
%\begin{figure}
%    \centering
%    \includegraphics[width=0.6\linewidth]{heat_pp_comp_noind_toi238b_upd.pdf}
%    \caption{Volume heating rates for TOI-238\,b: the H-He atmosphere with solar metallicity (top panel) and the atmosphere enriched in metals 200 times over solar metallicity (bottom panel). The plot format matches that of Fig.\,\ref{fig:heat_compn_noind}.}
%    \label{fig:heat_toi238b}
%\end{figure}
%
%\begin{figure}
%    \centering
%    \includegraphics[width=0.5\linewidth]{Hpops_pp_comp_noind_toi238b_upd.pdf}
%    \caption{Population of hydrogen species and heavy (metal) ions: the H-He atmosphere with solar metallicity (top panel) and the atmosphere enriched in metals 200 times over solar metallicity (bottom panel). The plot format matches that of Fig.\,\ref{fig:hpops_compn_noind}.}
%    \label{fig:hpops_toi238b}
%\end{figure}

\section{List of the simulations}\label{apx:tables}
%\label{appendix}
In Tables\,\ref{tab:gj9827d_simlist0} and \ref{tab:gj9827d_simlist1}, we provide the summary of the simulations performed for GJ\,9827\,d. Table\,\ref{tab:gj9827d_simlist0} includes the input parameters that can vary between the simulations (temperature $T_0$ and pressure $P_0$ adopted at the lower boundary of the simulation domain, orbital separation $a$, incoming X-ray and EUV fluxes $F_{\rm X}$ and $F_{\rm EUV}$, and the position of the upper boundary $R_1$ relative to the Roche radius $R_{\rm roche}$) and the final mass loss rates ($\dot{M}$). For the latter, values marked with asterisk correspond to the simulations where the exobase turns out to lie below the sonic point, hence the hydrodynamic approach is, strictly speaking, invalid. For these cases, we assumed that the escape occurs in the form of the enhanced Jeans escape and re-defined $\dot{M}$ using the simulations outputs below the exobase following \citet{Merry_Shiz1994P&SS...42..409M}. The parameters that are set uniformly for all simulations are planetary radius $R_{\rm pl} = 2.02$\,\Rer, mass $M_{\rm pl} = 3.53$\,\Mer, and stellar mass $M_{\rm *} = 0.606$\,\Msun.

Table\,\ref{tab:gj9827d_simlist1} includes some of the outputs reflecting the (thermo-)dynamical properties of the outflow: the peak temperature of the atmosphere $T_{\rm max}$, bulk velocity at the Roche radius $V_{\rm roche}$ (and Roche radius $R_{\rm roche}$ for the reference), sonic radius $R_{\rm S}$, position of the exobase $R_{\rm exo}$ and exospheric temperature $T_{\rm exo}$, indicative effective radius $R_{\rm eff}$(20 eV), and the actual position of the hydrogen ionisation front $R_{\rm ion}$ (as the region dominated by photoionisation is quite broad when one accounts for the realistic SED, it can be considered as the lowermost altitude where photioionisation becomes relevant). The value of $R_{\rm eff}$(20 eV) was defined as described in K18 (but using the full atmospheric opacity), assuming that the whole XUV flux is emitted at the single photon energy of 20\,eV and the only source of the heating is hydrogen photoionisation. From the difference between $R_{\rm eff}$ and $R_{\rm ion}$, one can see that though $R_{\rm eff}$ is a useful approximation allowing to obtain adequate mass loss estimates using the energy-limited equation, it does not reflect a real picture of photoionisation in planetary atmospheres.

The simulation ID's have the following format: \{atmosphere's type\}\_\{``v'' if adiabatic advection in Cloudy is ON\}\_\{IR enhancement flag\}\_\{additional information\}. Atmosphere's type are ``H-He-M'' - H-He atmospheres with solar metallicities (default), ``H-He'' - H and He with solar abundances without heavier elements, ``H'' - pure H atmospheres, ``H2O-z\{i\}'' - water-rich atmospheres with $f_{\rm H_2O}^{\rm m} = i\%$, and ``HMx\{j\}'' - metal-enriched atmospheres with $Z = j\times Z_{\odot}$. IR enhancement flag is ``LDD=0'' for the default case (no enhancement) and ``LDD=1'' for the IR-enhanced SED. Additional information can include: ``NOIND'' - no induced processes, ``NOLyaP'' - no Ly$\alpha$ pumping, ``300K''/``400K'' - planet at the orbit corresponding to $T_{\rm eq}$ of 300\,K or 400\,K, ``0.5*Lx''/``0.1*Lx'' - $F_{\rm X}$ reduced by the factor of 0.5 or 0.1, ``XUV-5Gyr''/``XUV-9Gyr'' - $F_{\rm X}$ and $F_{\rm EUV}$ reduced to the values corresponding to the star's ages of 5\,Gyr and 9\,Gyr (factors of $\sim4.4$ and $\sim7.7$ for X-ray, and of $\sim3.2$ and $\sim4.5$ for EUV), and ``SED-WASP43'' - uses the SED of WASP-43 \citep[as described in][]{Piaulet2024ApJ...974L..10P} instead of the SED of GJ-436 used in all other cases.
Here, we omit some of the less relevant simulations; the full tables can be found in the online materials along with the full simulation outputs and Cloudy inputs.

%input + mass loss
\begin{table*}[]
    \centering
    \caption{Simulation list for GJ\,9827\,d: variable input parameters (lower boundary temperature $T_{0}$ and pressure $P_{0}$, orbital separation $a$, total X-ray $F_{\rm X}$ and EUV $F_{\rm EUV}$ fluxed, and position of the upper boundary $R_1$ relative to the Roche radius) and mass loss rates.}
    \begin{tabular}{l|c|c|c|c|c|c|c}
    \toprule
    simulation ID & $T_{0}$ & $P_{0}$  & $a$    & $F_{\rm X}$          & $F_{\rm EUV}$ & $R_1/R_{\rm roche}$  & $\dot{M}$  \\ 
             & [K]   & [mbar] & [au] & [erg/s/${\rm cm^2}$] & [erg/s/${\rm cm^2}$]  &    & [$10^8$\,g/s]    \\ 
    \midrule
H-He-M\_LDD=0 & 611.1 &  0.5 & 0.055 & 1332.6 & 2962.9 & 1.50 & 374.34 \\ 
    \hline
H-He-M\_LDD=0\_0.1*Lx & 611.3 &  0.5 & 0.055 & 133.7 & 2926.6 & 1.50 & 348.26 \\ 
    \hline
H-He-M\_LDD=0\_300K & 300.1 &  0.5 & 0.249 & 64.8 & 144.1 & 0.50 & 0.65* \\ 
    \hline
%H-He-M\_LDD=0\_400K & 400.1 &  0.5 & 0.140 & 204.8 & 455.4 & 1.50 & 4.91* \\ 
%    \hline
H-He-M\_LDD=0\_NOIND & 609.1 &  0.5 & 0.055 & 1332.6 & 2962.9 & 1.50 & 126.88 \\ 
    \hline
H-He-M\_LDD=0\_NOLyaP & 609.2 &  0.5 & 0.055 & 1332.6 & 2962.9 & 1.50 & 131.65 \\ 
    \hline
%H-He-M\_LDD=0\_SED-WASP43 & 611.8 &  0.5 & 0.055 & 1352.7 & 2942.8 & 1.50 & 401.00 \\ 
    %\hline
H-He-M\_LDD=0\_XUV-9Gyr & 611.6 &  0.5 & 0.055 & 172.7 & 652.0 & 1.50 & 217.10 \\ 
    \hline
H-He-M\_LDD=1 & 611.2 &  0.5 & 0.055 & 1372.4 & 2923.2 & 1.50 & 147.71 \\ 
    \hline
H-He-M\_LDD=1\_0.1*Lx & 611.3 &  0.5 & 0.055 & 133.7 & 2926.6 & 1.50 & 142.14 \\ 
    \hline
H-He-M\_LDD=1\_0.5*Lx & 611.2 &  0.5 & 0.055 & 671.1 & 2938.1 & 1.50 & 145.00 \\ 
    \hline
H-He-M\_LDD=1\_300K & 300.0 &  0.5 & 0.249 & 64.8 & 144.1 & 0.50 & 0.62* \\ 
    \hline
%H-He-M\_LDD=1\_400K & 400.0 &  0.5 & 0.140 & 204.8 & 455.4 & 1.50 & 3.90* \\ 
%    \hline
H-He-M\_LDD=1\_NOIND & 611.1 &  0.5 & 0.055 & 1372.4 & 2923.2 & 1.50 & 121.85 \\ 
    \hline
H-He-M\_LDD=1\_NOLyaP & 611.2 &  0.5 & 0.055 & 1372.4 & 2923.2 & 1.50 & 123.35 \\ 
    \hline
%H-He-M\_LDD=1\_XUV-5Gyr & 611.5 &  0.5 & 0.055 & 301.9 & 939.0 & 1.50 & 57.24 \\ 
%    \hline
H-He-M\_LDD=1\_XUV-9Gyr & 611.6 &  0.5 & 0.055 & 172.7 & 652.0 & 1.50 & 20.11* \\ 
    \hline
H-He-M\_v\_LDD=1 & 610.0 & 10.0 & 0.055 & 1372.4 & 2923.2 & 1.50 & 158.95 \\ 
    \hline
H-He\_LDD=0\_NOIND & 609.0 &  0.5 & 0.055 & 1332.6 & 2962.9 & 1.50 & 130.53 \\ 
    \hline
H-He\_LDD=1 & 611.9 &  0.5 & 0.055 & 1372.4 & 2923.2 & 1.50 & 155.71 \\ 
    \hline
H-He\_LDD=1\_NOIND & 611.9 &  0.5 & 0.055 & 1372.4 & 2923.2 & 1.50 & 130.31 \\ 
    \hline
H2O-z50\_LDD=0 & 610.2 &  0.5 & 0.055 & 1332.6 & 2962.9 & 1.50 & 285.52 \\ 
    \hline
H2O-z50\_LDD=0\_0.14AU & 610.1 &  0.5 & 0.140 & 204.8 & 455.4 & 1.50 & 6.96* \\ 
    \hline
H2O-z50\_LDD=0\_300K & 300.3 &  0.1 & 0.249 & 64.8 & 144.1 & 0.40 & 0.04* \\ 
    \hline
%H2O-z50\_LDD=0\_400K & 400.9 &  0.1 & 0.140 & 204.8 & 455.4 & 1.00 & 2.95* \\ 
%    \hline
H2O-z50\_LDD=0\_NOIND & 609.4 &  0.5 & 0.055 & 1332.6 & 2962.9 & 1.50 & 97.59 \\ 
    \hline
H2O-z50\_LDD=1 & 611.4 &  0.5 & 0.055 & 1372.4 & 2923.2 & 1.50 & 117.92 \\ 
    \hline
H2O-z50\_LDD=1\_300K & 300.0 &  0.5 & 0.249 & 64.8 & 144.1 & 0.50 & 0.24* \\ 
    \hline
H2O-z50\_LDD=1\_NOIND & 611.4 &  0.5 & 0.055 & 1372.4 & 2923.2 & 1.50 & 100.23 \\ 
    \hline
H2O-z70\_LDD=0 & 610.8 &  0.5 & 0.055 & 1372.4 & 2923.2 & 1.50 & 173.54 \\ 
    \hline
H2O-z70\_LDD=0\_NOIND & 609.5 &  0.5 & 0.055 & 1332.6 & 2962.9 & 1.50 & 71.88 \\ 
    \hline
H2O-z70\_LDD=1 & 610.8 &  0.5 & 0.055 & 1372.4 & 2923.2 & 1.50 & 85.46 \\ 
    \hline
H2O-z70\_LDD=1\_NOIND & 610.8 &  0.5 & 0.055 & 1372.4 & 2923.2 & 1.50 & 76.81 \\ 
    \hline
H2O-z70\_LDD=1 & 610.4 & 10.0 & 0.055 & 1372.4 & 2923.2 & 1.50 & 140.92 \\ 
    \hline
H2O-z80\_LDD=0\_NOIND & 609.6 &  0.5 & 0.055 & 1332.6 & 2962.9 & 1.50 & 64.54 \\ 
    \hline
H2O-z80\_LDD=1 & 610.8 &  0.5 & 0.055 & 1372.4 & 2923.2 & 1.50 & 75.81 \\ 
    \hline
H2O-z80\_LDD=1\_NOIND & 610.7 &  0.5 & 0.055 & 1372.4 & 2923.2 & 1.50 & 70.17 \\ 
    \hline
H2O-z90\_LDD=0 & 610.9 &  0.5 & 0.055 & 1372.4 & 2923.2 & 1.50 & 101.48 \\ 
    \hline
H2O-z90\_LDD=0\_NOIND & 609.7 &  0.5 & 0.055 & 1332.6 & 2962.9 & 1.50 & 62.83 \\ 
    \hline
H2O-z90\_LDD=1 & 610.9 &  0.5 & 0.055 & 1372.4 & 2923.2 & 1.50 & 72.64 \\ 
    \hline
HMx010\_LDD=0 & 612.0 &  0.5 & 0.055 & 1372.4 & 2923.2 & 1.50 & 584.25 \\ 
    \hline
HMx010\_LDD=0\_NOIND & 608.0 &  0.5 & 0.055 & 1332.6 & 2962.9 & 1.50 & 194.57 \\ 
    \hline
HMx010\_LDD=1 & 612.0 &  0.5 & 0.055 & 1372.4 & 2923.2 & 1.50 & 184.69 \\ 
    \hline
HMx010\_LDD=1\_NOIND & 612.0 &  0.5 & 0.055 & 1372.4 & 2923.2 & 1.50 & 135.61 \\ 
    \hline
HMx050\_LDD=1 & 612.1 & 10.0 & 0.055 & 1372.4 & 2923.2 & 1.50 & 198.53 \\ 
    \hline
HMx100\_LDD=0 & 612.2 &  0.5 & 0.055 & 1372.4 & 2923.2 & 1.50 & 329.18 \\ 
    \hline
HMx100\_LDD=0\_NOIND & 608.4 &  0.5 & 0.055 & 1332.6 & 2962.9 & 1.50 & 178.58 \\ 
    \hline
HMx100\_LDD=1 & 612.2 &  0.5 & 0.055 & 1372.4 & 2923.2 & 1.50 & 121.43 \\ 
    \hline
HMx200\_LDD=0 & 612.1 &  0.5 & 0.055 & 1372.4 & 2923.2 & 1.50 & 202.17 \\ 
    \hline
HMx200\_LDD=0\_NOIND & 608.5 &  0.5 & 0.055 & 1332.6 & 2962.9 & 1.50 & 148.09 \\ 
    \hline
HMx200\_LDD=1 & 612.1 &  0.5 & 0.055 & 1372.4 & 2923.2 & 1.50 & 109.59 \\ 
    \hline
HMx200\_LDD=1\_NOIND & 612.5 &  0.5 & 0.055 & 1372.4 & 2923.2 & 1.50 & 89.76 \\ 
    \hline
HMx300\_LDD=0 & 612.4 &  0.5 & 0.055 & 1372.4 & 2923.2 & 1.50 & 145.27 \\ 
    \hline
HMx300\_LDD=0\_NOIND\_tmp & 608.7 &  0.5 & 0.055 & 1332.6 & 2962.9 & 1.50 & 134.71 \\ 
    \hline
HMx300\_LDD=1 & 612.2 &  0.5 & 0.055 & 1372.4 & 2923.2 & 1.50 & 102.74 \\ 
    \hline
HMx300\_LDD=1\_NOIND & 612.7 &  0.5 & 0.055 & 1372.4 & 2923.2 & 1.50 & 83.86 \\ 
    \hline
HMx500\_LDD=0 & 612.3 &  0.5 & 0.055 & 1372.4 & 2923.2 & 1.00 & 143.57 \\ 
    \hline
HMx500\_LDD=0\_NOIND & 608.8 &  0.5 & 0.055 & 1332.6 & 2962.9 & 1.00 & 123.71 \\ 
    \hline
HMx500\_LDD=1 & 612.3 &  0.5 & 0.055 & 1372.4 & 2923.2 & 1.00 & 117.61 \\ 
    \hline
%HMx800\_LDD=0\_NOIND & 608.9 &  0.5 & 0.055 & 1332.6 & 2962.9 & 1.20 & 153.12 \\ 
%    \hline
%HMx800\_LDD=1 & 612.6 &  0.5 & 0.055 & 1372.4 & 2923.2 & 1.20 & 126.64 \\ 
%    \hline
H\_LDD=0 & 611.9 &  0.5 & 0.055 & 1372.4 & 2923.2 & 1.50 & 592.17 \\ 
    \hline
H\_LDD=1 & 611.9 &  0.5 & 0.055 & 1372.4 & 2923.2 & 1.50 & 236.99 \\ 
    \bottomrule 
    \end{tabular}
    \\
    \footnotesize{*atmospheric escape occurs in Jeans-like regime.}
    \label{tab:gj9827d_simlist0}
\end{table*}

%outputs
\begin{table*}[]
    \centering
    \caption{Output parameters for GJ\,9827\,d: peak temperature $T_{\rm max}$, bulk outflow velocity at the Roche radius $V_{\rm roche}$, position of the Roche radius $R_{\rm roche}$, sonic radius $R_{\rm S}$, exobase position $R_{\rm exo}$, temperature at the exobase $T_{\rm exo}$, effective radius $R_{\rm eff}$, and the position of the maximum of $n_{\rm H^+}$ -- $R_{\rm ion}$.}
    \begin{tabular}{l|c|c|c|c|c|c|c|c}
    \toprule
    simulation ID & $T_{\rm max}$ & $V_{\rm roche}$ & $R_{\rm roche}$ & $R_{\rm S}$ & $R_{\rm exo}$ & $T_{\rm exo}$ & $R_{\rm eff}$(20 eV)  & $R_{\rm ion}$  \\ 
             & [K]            & [km/s]           & [$R_{\rm pl}$]  & [$R_{\rm pl}$] & [$R_{\rm pl}$] & [K]        &  [$R_{\rm pl}$]       & [$R_{\rm pl}$]    \\ 
    \midrule
H-He-M\_LDD=0 & 1315.1 & 6.4 & 11.5 & 5.4 & 11.8 & 962.1 & 8.5 & 1.4 \\ 
    \hline
H-He-M\_LDD=0\_0.1*Lx & 1328.3 & 6.3 & 11.5 & 5.3 & 11.6 & 957.7 & 8.3 & 1.4 \\ 
    \hline
H-He-M\_LDD=0\_300K & 1504.8 & 2.0 & 52.1 & 16.1 & 3.4 & 946.2 & 2.4 & 1.2 \\ 
    \hline
%H-He-M\_LDD=0\_400K & 1443.5 & 4.3 & 29.3 & 9.8 & 4.8 & 782.7 & 4.1 & 1.3 \\ 
%    \hline
H-He-M\_LDD=0\_NOIND & 1396.7 & 6.3 & 11.5 & 5.0 & 7.9 & 921.9 & 5.7 & 1.4 \\ 
    \hline
H-He-M\_LDD=0\_NOLyaP & 1388.9 & 6.3 & 11.5 & 5.0 & 7.9 & 919.4 & 5.8 & 1.4 \\ 
    \hline
%H-He-M\_LDD=0\_SED-WASP43 & 1349.3 & 6.7 & 11.5 & 5.2 & 11.6 & 890.6 & 7.8 & 1.4 \\ 
%    \hline
H-He-M\_LDD=0\_XUV-9Gyr & 1571.4 & 4.1 & 11.5 & 7.5 & 9.6 & 567.9 & 9.0 & 1.4 \\ 
    \hline
H-He-M\_LDD=1 & 1421.9 & 6.8 & 11.5 & 4.5 & 8.1 & 924.4 & 5.4 & 1.4 \\ 
    \hline
H-He-M\_LDD=1\_0.1*Lx & 1426.1 & 6.8 & 11.5 & 4.5 & 8.0 & 922.9 & 5.4 & 1.4 \\ 
    \hline
H-He-M\_LDD=1\_0.5*Lx & 1429.2 & 6.8 & 11.5 & 4.5 & 8.0 & 925.4 & 5.4 & 1.4 \\ 
    \hline
H-He-M\_LDD=1\_300K & 1523.0 & 2.2 & 52.1 & 14.2 & 3.3 & 961.7 & 2.3 & 1.2 \\ 
    \hline
%H-He-M\_LDD=1\_400K & 1488.5 & 4.5 & 29.3 & 8.7 & 4.4 & 817.7 & 3.4 & 1.3 \\ 
%    \hline
H-He-M\_LDD=1\_NOIND & 1474.4 & 6.8 & 11.5 & 4.4 & 7.5 & 925.8 & 5.0 & 1.6 \\ 
    \hline
H-He-M\_LDD=1\_NOLyaP & 1465.1 & 6.8 & 11.5 & 4.4 & 7.5 & 911.5 & 5.1 & 1.6 \\ 
    \hline
%H-He-M\_LDD=1\_XUV-5Gyr & 1218.8 & 4.9 & 11.5 & 6.0 & 6.3 & 736.1 & 5.0 & 1.5 \\ 
%    \hline
H-He-M\_LDD=1\_XUV-9Gyr & 1179.7 & 4.3 & 11.5 & 6.5 & 5.9 & 690.1 & 4.9 & 1.5 \\ 
    \hline
H-He-M\_v\_LDD=1 & 1706.6 & 7.3 & 11.5 & 5.2 & 9.5 & 1432.4 & 6.5 & 1.6 \\ 
    \hline
H-He\_LDD=0\_NOIND & 1379.9 & 6.3 & 11.5 & 5.0 & 8.0 & 915.5 & 5.8 & 1.4 \\ 
    \hline
H-He\_LDD=1 & 1400.6 & 6.8 & 11.5 & 4.5 & 8.2 & 920.7 & 5.6 & 1.4 \\ 
    \hline
H-He\_LDD=1\_NOIND & 1443.7 & 6.7 & 11.5 & 4.5 & 7.7 & 922.0 & 5.2 & 1.4 \\ 
    \hline
H2O-z50\_LDD=0 & 2249.7 & 11.9 & 11.5 & 3.5 & 10.2 & 1606.4 & 4.9 & 1.3 \\ 
    \hline
H2O-z50\_LDD=0\_0.14AU & 1695.7 & 11.5 & 29.3 & 4.1 & 3.8 & 1430.0 & 2.3 & 1.2 \\ 
    \hline
H2O-z50\_LDD=0\_300K & 1825.8 & 8.4 & 52.1 & 2.8 & 2.2 & 966.8 & 1.5 & 1.1 \\ 
    \hline
%H2O-z50\_LDD=0\_400K & 1867.8 & 10.6 & 29.3 & 3.4 & 3.0 & 1356.0 & 1.9 & 1.1 \\ 
%    \hline
H2O-z50\_LDD=0\_NOIND & 1984.9 & 11.8 & 11.5 & 3.0 & 6.4 & 1502.4 & 3.1 & 1.3 \\ 
    \hline
H2O-z50\_LDD=1 & 2014.1 & 11.9 & 11.5 & 2.8 & 6.6 & 1386.3 & 3.0 & 1.4 \\ 
    \hline
H2O-z50\_LDD=1\_300K & 1779.3 & 8.6 & 52.1 & 3.1 & 2.2 & 1100.7 & 1.5 & 1.2 \\ 
    \hline
H2O-z50\_LDD=1\_NOIND & 1999.8 & 11.8 & 11.5 & 2.7 & 6.2 & 1377.6 & 2.9 & 1.3 \\ 
    \hline
H2O-z70\_LDD=0 & 2774.5 & 14.3 & 11.5 & 2.6 & 7.6 & 1633.0 & 3.0 & 1.2 \\ 
    \hline
H2O-z70\_LDD=0\_NOIND & 2505.4 & 13.3 & 11.5 & 2.3 & 5.3 & 1547.6 & 2.2 & 1.2 \\ 
    \hline
H2O-z70\_LDD=1 & 2473.3 & 13.1 & 11.5 & 2.2 & 5.5 & 1417.1 & 2.2 & 1.2 \\ 
    \hline
H2O-z70\_LDD=1\_NOIND & 2454.5 & 13.0 & 11.5 & 2.2 & 5.2 & 1411.6 & 2.1 & 1.2 \\ 
    \hline
H2O-z70\_LDD=1 & 2675.5 & 13.7 & 11.5 & 2.3 & 6.3 & 1441.1 & 2.6 & 1.3 \\ 
    \hline
H2O-z80\_LDD=0\_NOIND & 2817.9 & 13.7 & 11.5 & 2.1 & 4.8 & 1508.7 & 1.9 & 1.1 \\ 
    \hline
H2O-z80\_LDD=1 & 2753.4 & 13.4 & 11.5 & 2.0 & 5.0 & 1389.0 & 1.9 & 1.2 \\ 
    \hline
H2O-z80\_LDD=1\_NOIND & 2731.2 & 13.3 & 11.5 & 2.0 & 4.9 & 1386.1 & 1.8 & 1.2 \\ 
    \hline
H2O-z90\_LDD=0 & 3426.0 & 14.7 & 11.5 & 1.9 & 5.4 & 1424.2 & 1.9 & 1.1 \\ 
    \hline
H2O-z90\_LDD=0\_NOIND & 3123.0 & 13.7 & 11.5 & 1.8 & 4.5 & 1384.3 & 1.6 & 1.1 \\ 
    \hline
H2O-z90\_LDD=1 & 3019.3 & 13.4 & 11.5 & 1.8 & 4.7 & 1286.7 & 1.7 & 1.1 \\ 
    \hline
HMx010\_LDD=0 & 1739.1 & 8.5 & 11.5 & 5.0 & 14.5 & 1382.7 & 8.7 & 1.7 \\ 
    \hline
HMx010\_LDD=0\_NOIND & 1489.4 & 8.7 & 11.5 & 4.2 & 9.2 & 1307.8 & 5.6 & 1.8 \\ 
    \hline
HMx010\_LDD=1 & 1541.4 & 9.3 & 11.5 & 3.6 & 8.3 & 1193.6 & 4.9 & 1.9 \\ 
    \hline
HMx010\_LDD=1\_NOIND & 1514.2 & 9.3 & 11.5 & 3.4 & 7.4 & 1174.2 & 4.3 & 1.6 \\ 
    \hline
HMx050\_LDD=1 & 2108.6 & 11.8 & 11.5 & 3.1 & 8.4 & 1445.1 & 4.0 & 1.8 \\ 
    \hline
HMx100\_LDD=0 & 2955.2 & 14.1 & 11.5 & 3.0 & 10.4 & 1764.4 & 4.4 & 2.0 \\ 
    \hline
HMx100\_LDD=0\_NOIND & 2724.9 & 13.7 & 11.5 & 2.8 & 8.3 & 1728.7 & 3.5 & 1.7 \\ 
    \hline
HMx100\_LDD=1 & 2534.5 & 13.3 & 11.5 & 2.5 & 6.5 & 1506.1 & 2.7 & 1.5 \\ 
    \hline
HMx200\_LDD=0 & 3693.4 & 16.1 & 11.5 & 2.3 & 7.8 & 1705.5 & 2.8 & 1.6 \\ 
    \hline
HMx200\_LDD=0\_NOIND & 3503.0 & 15.6 & 11.5 & 2.2 & 6.8 & 1672.4 & 2.5 & 1.5 \\ 
    \hline
HMx200\_LDD=1 & 3171.5 & 14.6 & 11.5 & 2.1 & 5.9 & 1499.3 & 2.1 & 1.3 \\ 
    \hline
HMx200\_LDD=1\_NOIND & 3069.0 & 14.1 & 11.5 & 2.0 & 5.5 & 1486.1 & 2.0 & 1.3 \\ 
    \hline
HMx300\_LDD=0 & 4104.6 & 16.2 & 11.5 & 1.9 & 6.3 & 1568.5 & 2.2 & 1.4 \\ 
    \hline
HMx300\_LDD=0\_NOIND\_tmp & 4035.7 & 15.1 & 11.5 & 2.1 & 6.2 & 1574.0 & 2.1 & 1.4 \\ 
    \hline
HMx300\_LDD=1 & 3514.7 & 16.5 & 11.5 & 1.9 & 4.9 & 1321.2 & 1.9 & 1.2 \\ 
    \hline
HMx300\_LDD=1\_NOIND & 3358.7 & 14.4 & 11.5 & 1.9 & 5.1 & 1416.2 & 1.8 & 1.2 \\ 
    \hline
HMx500\_LDD=0 & 4412.2 & 16.1 & 11.5 & 1.8 & 6.0 & 1407.1 & 1.9 & 1.2 \\ 
    \hline
HMx500\_LDD=0\_NOIND & 4231.0 & 15.7 & 11.5 & 1.8 & 5.6 & 1379.7 & 1.8 & 1.2 \\ 
    \hline
HMx500\_LDD=1 & 3871.6 & 14.8 & 11.5 & 1.7 & 5.7 & 1313.8 & 1.8 & 1.2 \\ 
    \hline
HMx800\_LDD=0\_NOIND & 4135.9 & 13.7 & 11.5 & 1.7 & 5.6 & 994.0 & 1.8 & 1.1 \\ 
    \hline
%HMx800\_LDD=1 & 3518.5 & 12.3 & 11.5 & 1.6 & 5.4 & 954.3 & 1.7 & 1.1 \\ 
%    \hline
H\_LDD=0 & 1054.8 & 4.7 & 11.5 & 6.3 & 13.8 & 720.3 & 10.8 & 1.5 \\ 
    \hline
H\_LDD=1 & 961.8 & 5.5 & 11.5 & 5.3 & 9.6 & 722.6 & 7.3 & 1.5 \\ 
    \bottomrule 
    \end{tabular}

    \label{tab:gj9827d_simlist1}
\end{table*}

%% toi 238 b
\begin{table*}[]
    \centering
    \caption{Simulation list for TOI-238\,b: variable input parameters (lower boundary temperature $T_{0}$ and pressure $P_{0}$, orbital separation $a$, total X-ray $F_{\rm X}$ and EUV $F_{\rm EUV}$ fluxed, and position of the upper boundary $R_1$ relative to the Roche radius) and mass loss rates.}
    \begin{tabular}{l|c|c|c|c|c|c|c}
    simulation ID & $T_{0}$ & $P_{0}$  & $a$    & $F_{\rm X}$          & $F_{\rm EUV}$ & $R_1/R_{\rm roche}$  & $\dot{M}$  \\ 
             & [K]   & [mbar] & [au] & [erg/s/${\rm cm^2}$] & [erg/s/${\rm cm^2}$]  &  & [$10^8$\,g/s]    \\ 
    \hline \hline
H-He-M\_LDD=0 & 1270.0 &  1.0 & 0.026 & 6680.4 & 13835.9 & 1.80 & 6486.21 \\ 
    \hline
H-He-M\_LDD=0\_NOIND & 1271.0 &  1.0 & 0.026 & 6680.4 & 13835.9 & 1.80 & 705.76 \\ 
    \hline
H-He-M\_LDD=1 & 1269.0 &  1.0 & 0.026 & 6680.4 & 13835.9 & 1.80 & 1983.36 \\ 
    \hline
H-He-M\_LDD=1\_NOIND & 1268.0 &  1.0 & 0.026 & 6680.4 & 13835.9 & 1.80 & 577.93 \\ 
    \hline
H2O-z10\_LDD=0 & 1269.1 &  1.0 & 0.026 & 6680.4 & 13835.9 & 1.80 & 11040.07 \\ 
    \hline
H2O-z10\_LDD=0\_NOIND & 1271.1 &  1.0 & 0.026 & 6680.4 & 13835.9 & 1.80 & 869.76 \\ 
    \hline
H2O-z10\_LDD=1 & 1269.1 &  1.0 & 0.026 & 6680.4 & 13835.9 & 1.80 & 2822.62 \\ 
    \hline
H2O-z10\_LDD=1\_NOIND & 1268.1 &  1.0 & 0.026 & 6680.4 & 13835.9 & 1.80 & 757.07 \\ 
    \hline
H2O-z30\_LDD=1 & 1269.2 &  1.0 & 0.026 & 6680.4 & 13835.9 & 1.80 & 2403.83 \\ 
    \hline
H2O-z50\_LDD=0 & 1270.3 &  1.0 & 0.026 & 6680.4 & 13835.9 & 1.80 & 5924.78 \\ 
    \hline
H2O-z50\_LDD=0\_NOIND & 1271.2 &  1.0 & 0.026 & 6680.4 & 13835.9 & 1.80 & 505.39 \\ 
    \hline
H2O-z50\_LDD=1 & 1269.3 &  1.0 & 0.026 & 6680.4 & 13835.9 & 1.80 & 1385.97 \\ 
    \hline
H2O-z50\_LDD=1\_NOIND & 1268.2 &  1.0 & 0.026 & 6680.4 & 13835.9 & 1.80 & 378.43 \\ 
    \hline
H2O-z70\_LDD=0 & 1270.4 &  1.0 & 0.026 & 6680.4 & 13835.9 & 1.80 & 4405.03 \\ 
    \hline
H2O-z70\_LDD=0\_NOIND & 1271.3 &  1.0 & 0.026 & 6680.4 & 13835.9 & 1.80 & 390.87 \\ 
    \hline
H2O-z70\_LDD=1 & 1269.4 &  1.0 & 0.026 & 6680.4 & 13835.9 & 1.80 & 1010.20 \\ 
    \hline
H2O-z70\_LDD=1\_NOIND & 1268.3 &  1.0 & 0.026 & 6680.4 & 13835.9 & 1.80 & 296.39 \\ 
    \hline
H2O-z80\_LDD=0 & 1269.5 &  1.0 & 0.026 & 6680.4 & 13835.9 & 1.80 & 3907.16 \\ 
    \hline
H2O-z80\_LDD=1 & 1269.5 &  1.0 & 0.026 & 6680.4 & 13835.9 & 1.80 & 872.58 \\ 
    \hline
H2O-z90\_LDD=0 & 1269.6 &  1.0 & 0.026 & 6680.4 & 13835.9 & 1.80 & 3465.7 \\ 
    \hline
H2O-z90\_LDD=0\_NOIND & 1296.6 &  0.1 & 0.026 & 6680.4 & 13835.9 & 1.80 & 158.71 \\ 
    \hline
H2O-z90\_LDD=1 & 1294.8 &  0.1 & 0.026 & 6680.4 & 13835.9 & 1.80 & 431.62 \\ 
    \hline
H2O-z90\_LDD=1 & 1269.6 &  1.0 & 0.026 & 6680.4 & 13835.9 & 1.80 & 722.67 \\ 
    \hline
H2O-z95\_LDD=0 & 1292.1 &  1.0 & 0.026 & 6680.4 & 13835.9 & 1.80 & 2610.4 \\ 
    \hline
H2O-z95\_LDD=1 & 1292.1 &  0.1 & 0.026 & 6680.4 & 13835.9 & 1.80 & 219.43 \\ 
    \hline
HMx010\_LDD=0\_NOIND & 1271.5 &  0.5 & 0.026 & 6680.4 & 13835.9 & 1.80 & 1474.49 \\ 
    \hline
HMx010\_LDD=1 & 1269.0 &  0.5 & 0.026 & 6680.4 & 13835.9 & 1.80 & 8753.04 \\ 
    \hline
HMx030\_LDD=0 & 1270.1 &  0.5 & 0.026 & 6680.4 & 13835.9 & 1.80 & 23687.40 \\ 
    \hline
HMx030\_LDD=0\_NOIND & 1271.6 &  0.5 & 0.026 & 6680.4 & 13835.9 & 1.80 & 949.86 \\ 
    \hline
HMx030\_LDD=1 & 1269.1 &  0.5 & 0.026 & 6680.4 & 13835.9 & 1.80 & 13085.04 \\ 
    \hline
HMx030\_LDD=1\_NOIND & 1268.6 &  0.5 & 0.026 & 6680.4 & 13835.9 & 1.80 & 511.20 \\ 
    \hline
HMx050\_LDD=1 & 1269.2 &  0.5 & 0.026 & 6680.4 & 13835.9 & 1.80 & 11113.71 \\ 
    \hline
HMx050\_LDD=1\_NOIND & 1268.7 &  0.5 & 0.026 & 6680.4 & 13835.9 & 1.80 & 462.27 \\ 
    \hline
HMx100\_LDD=0 & 1269.3 &  0.5 & 0.026 & 6680.4 & 13835.9 & 1.80 & 11842.94 \\ 
    \hline
HMx100\_LDD=0\_NOIND & 1271.8 &  0.5 & 0.026 & 6680.4 & 13835.9 & 1.80 & 593.68 \\ 
    \hline
HMx100\_LDD=1 & 1269.3 &  0.5 & 0.026 & 6680.4 & 13835.9 & 1.80 & 3513.11 \\ 
    \hline
HMx100\_LDD=1\_NOIND & 1268.8 &  0.5 & 0.026 & 6680.4 & 13835.9 & 1.80 & 394.48 \\ 
    \hline
HMx200\_LDD=0 & 1269.4 &  0.5 & 0.026 & 6680.4 & 13835.9 & 1.80 & 1839.98 \\ 
    \hline
HMx200\_LDD=0\_NOIND & 1271.9 &  0.5 & 0.026 & 6680.4 & 13835.9 & 1.80 & 410.17 \\ 
    \hline
HMx200\_LDD=1 & 1269.4 &  0.5 & 0.026 & 6680.4 & 13835.9 & 1.80 & 1061.73 \\ 
    \hline
HMx300\_LDD=1 & 1269.5 &  0.5 & 0.026 & 6680.4 & 13835.9 & 1.80 & 1400.2 \\ 
    \hline
HMx300\_LDD=1 & 1269.5 &  0.5 & 0.026 & 6680.4 & 13835.9 & 1.80 & 688.37 \\ 
    \hline
HMx500\_LDD=0 & 1269.6 &  0.5 & 0.026 & 6680.4 & 13835.9 & 1.80 & 1000.7 \\ 
    \hline
HMx500\_LDD=1 & 1269.6 &  0.5 & 0.026 & 6680.4 & 13835.9 & 1.80 & 459.03 \\ 
    \hline
HMx600\_LDD=0 & 1269.7 &  0.5 & 0.026 & 6680.4 & 13835.9 & 1.80 & 386.43 \\ 
    \hline
    \hline 
    \end{tabular}

    \label{tab:toi238b_simlist0}
\end{table*}

\begin{table*}[]
    \centering
    \caption{Output parameters for TOI-238\,b: peak temperature $T_{\rm max}$, bulk outflow velocity at the Roche radius $V_{\rm roche}$, position of the Roche radius $R_{\rm roche}$, sonic radius $R_{\rm S}$, exobase position $R_{\rm exo}$, temperature at the exobase $T_{\rm exo}$, effective radius $R_{\rm eff}$, and the location of the maximum of $n_{\rm H^+}$ -- $R_{\rm ion}$. ``--'' in the exobase parameters means that the atmosphere is fully collisional within the simulation domain (exobase lies above the upper boundary).}
    \begin{tabular}{l|c|c|c|c|c|c|c|c}
    simulation ID & $T_{\rm max}$ & $V_{\rm roche}$ & $R_{\rm roche}$ & $R_{\rm S}$ & $R_{\rm exo}$ & $T_{\rm exo}$ & $R_{\rm eff}$(20 eV)  & $R_{\rm ion}$  \\ 
             & [K]            & [km/s]           & [$R_{\rm pl}$]  & [$R_{\rm pl}$] & [$R_{\rm pl}$] & [K]        &  [$R_{\rm pl}$]       & [$R_{\rm pl}$]    \\ 
    \hline \hline
H-He-M\_LDD=0 & 1922.0 & 7.1 & 6.2 & 3.3 & -- & -- & 10.3 & 1.8 \\ 
    \hline
H-He-M\_LDD=0\_NOIND & 2098.2 & 9.2 & 6.2 & 3.6 & -- & -- & 6.0 & 1.8 \\ 
    \hline
H-He-M\_LDD=1 & 2289.7 & 7.8 & 6.2 & 4.7 & -- & -- & 7.6 & 1.9 \\ 
    \hline
H-He-M\_LDD=1\_NOIND & 2074.9 & 10.3 & 6.2 & 3.0 & 10.3 & 912.1 & 4.7 & 2.7 \\ 
    \hline
H2O-z10\_LDD=0 & 2062.5 & 7.2 & 6.2 & 3.4 & -- & -- & 10.8 & 2.2 \\ 
    \hline
H2O-z10\_LDD=0\_NOIND & 2144.7 & 8.8 & 6.2 & 3.9 & -- & -- & 6.5 & 1.6 \\ 
    \hline
H2O-z10\_LDD=1 & 2472.8 & 5.9 & 6.2 & 5.4 & -- & -- & 8.5 & 2.2 \\ 
    \hline
H2O-z10\_LDD=1\_NOIND & 2110.3 & 10.1 & 6.2 & 3.3 & -- & -- & 5.1 & 3.2 \\ 
    \hline
H2O-z30\_LDD=1 & 2880.2 & 8.8 & 6.2 & 4.9 & -- & -- & 7.5 & 2.0 \\ 
    \hline
H2O-z50\_LDD=0 & 3016.8 & 8.6 & 6.2 & 3.2 & -- & -- & 9.7 & 1.5 \\ 
    \hline
H2O-z50\_LDD=0\_NOIND & 2932.2 & 12.6 & 6.2 & 2.7 & 10.8 & 1419.0 & 4.0 & 1.4 \\ 
    \hline
H2O-z50\_LDD=1 & 3324.8 & 12.9 & 6.2 & 3.4 & -- & -- & 5.4 & 1.6 \\ 
    \hline
H2O-z50\_LDD=1\_NOIND & 2634.4 & 12.6 & 6.2 & 2.2 & 9.2 & 1221.8 & 3.2 & 1.4 \\ 
    \hline
H2O-z70\_LDD=0 & 4320.5 & 10.7 & 6.2 & 3.2 & -- & -- & 8.3 & 1.3 \\ 
    \hline
H2O-z70\_LDD=0\_NOIND & 3554.9 & 14.6 & 6.2 & 2.1 & 9.6 & 1530.0 & 3.0 & 1.2 \\ 
    \hline
H2O-z70\_LDD=1 & 4144.0 & 15.6 & 6.2 & 2.5 & -- & -- & 3.9 & 1.3 \\ 
    \hline
H2O-z70\_LDD=1\_NOIND & 3102.0 & 13.7 & 6.2 & 1.9 & 8.2 & 1337.7 & 2.5 & 1.2 \\ 
    \hline
H2O-z80\_LDD=0 & 4864.9 & 12.1 & 6.2 & 3.0 & -- & -- & 4.9 & 1.2 \\ 
    \hline
H2O-z80\_LDD=1 & 4856.6 & 17.1 & 6.2 & 2.0 & -- & -- & 3.3 & 1.2 \\ 
    \hline
H2O-z90\_LDD=0 & 5484.9 & 13.3 & 6.2 & 2.2 & -- & -- & 3.6 & 1.1 \\ 
    \hline
H2O-z90\_LDD=0\_NOIND & 3598.6 & 14.4 & 6.2 & 1.6 & 6.5 & 1489.5 & 1.7 & 1.0 \\ 
    \hline
H2O-z90\_LDD=1 & 4500.7 & 15.8 & 6.2 & 1.6 & 9.6 & 1483.8 & 2.3 & 1.0 \\ 
    \hline
H2O-z90\_LDD=1 & 5446.5 & 17.4 & 6.2 & 1.7 & -- & -- & 2.7 & 1.1 \\ 
    \hline
H2O-z95\_LDD=0 & 5202.3 & 14.7 & 6.2 & 1.8 & 10.6 & 1314.0 & 3.0 & 1.1 \\ 
    \hline
H2O-z95\_LDD=1 & 3517.0 & 13.4 & 6.2 & 1.6 & 7.3 & 1302.2 & 1.8 & 1.0 \\ 
    \hline
HMx010\_LDD=0\_NOIND & 2536.8 & 8.6 & 6.2 & 4.4 & -- & -- & 7.5 & 3.6 \\ 
    \hline
HMx010\_LDD=1 & 3106.2 & 3.4 & 6.2 & 5.9 & -- & -- & 10.6 & 2.9 \\ 
    \hline
HMx030\_LDD=0 & 2820.6 & 8.6 & 6.2 & 3.8 & -- & -- & 11.0 & 2.8 \\ 
    \hline
HMx030\_LDD=0\_NOIND & 2852.4 & 11.2 & 6.2 & 3.4 & -- & -- & 5.9 & 2.8 \\ 
    \hline
HMx030\_LDD=1 & 3963.6 & 4.6 & 6.2 & 5.0 & -- & -- & 10.8 & 2.9 \\ 
    \hline
HMx030\_LDD=1\_NOIND & 2533.2 & 11.9 & 6.2 & 2.5 & 10.0 & 1086.6 & 3.9 & 2.2 \\ 
    \hline
HMx050\_LDD=1 & 4408.9 & 4.2 & 6.2 & 5.4 & -- & -- & 10.4 & 2.8 \\ 
    \hline
HMx050\_LDD=1\_NOIND & 2777.0 & 12.6 & 6.2 & 2.3 & 9.7 & 1172.0 & 3.5 & 2.0 \\ 
    \hline
HMx100\_LDD=0 & 3834.2 & 10.7 & 6.2 & 3.2 & -- & -- & 10.4 & 1.9 \\ 
    \hline
HMx100\_LDD=0\_NOIND & 3780.7 & 14.5 & 6.2 & 2.3 & -- & -- & 3.8 & 2.0 \\ 
    \hline
HMx100\_LDD=1 & 5036.3 & 14.3 & 6.2 & 4.1 & -- & -- & 6.9 & 2.2 \\ 
    \hline
HMx100\_LDD=1\_NOIND & 3262.8 & 13.7 & 6.2 & 2.0 & 9.3 & 1322.0 & 2.9 & 1.7 \\ 
    \hline
HMx200\_LDD=0 & 4575.4 & 15.8 & 6.2 & 2.8 & -- & -- & 5.3 & 1.4 \\ 
    \hline
HMx200\_LDD=0\_NOIND & 4407.2 & 15.9 & 6.2 & 1.8 & 9.7 & 1640.0 & 2.7 & 1.5 \\ 
    \hline
HMx200\_LDD=1 & 5346.8 & 17.6 & 6.2 & 2.1 & -- & -- & 3.6 & 2.0 \\ 
    \hline
HMx300\_LDD=0 & 5022.5 & 17.4 & 6.2 & 2.4 & -- & -- & 4.0 & 1.1 \\ 
    \hline
HMx300\_LDD=1 & 5521.1 & 17.2 & 6.2 & 1.7 & -- & -- & 2.8 & 1.6 \\ 
    \hline
HMx500\_LDD=0 & 5122.8 & 16.8 & 6.2 & 2.2 & -- & -- & 2.8 & 1.1 \\ 
    \hline
HMx500\_LDD=1 & 5319.3 & 16.2 & 6.2 & 1.5 & 10.1 & 1659.4 & 2.1 & 1.3 \\ 
    \hline
HMx600\_LDD=0 & 5026.4 & 15.4 & 6.2 & 1.5 & 9.5 & 1582.6 & 2.0 & 1.3 \\ 
    \hline
    \hline 
    \end{tabular}

    \label{tab:toi238b_simlist1}
\end{table*}

\end{appendix}

\end{document}